# Destruction and Resurrection of Atomic Giant resonances in Endohedral Atoms A@C$_{60}$

M. Ya. Amusia[1,2], A. S. Baltenkov[3], L. V. Chernysheva[2]

[1]Racah Institute of Physics, the Hebrew University, 91904 Jerusalem, Israel
[2]Ioffe Physical-Technical Institute, 194021 St. Petersburg, Russia
[3] Arifov's Institute of Electronics, Akademgorodok, 700125 Tashkent, Uzbekistan

**Abstract**

It is demonstrated that in photoabsorption by endohedral atoms some atomic Giant resonances are almost completely destroyed while the others are totally preserved due to different action on it of the fullerenes shell. As the first example we discuss the $4d^{10}$ Giant resonance in Xe@C$_{60}$ whereas as the second serves the Giant *autoionization* resonance in Eu@C$_{60}$.

The qualitative difference comes from the fact that photoelectrons from the $4d$ Giant resonance has small energies (tens of eV) and are strongly reflected by the C$_{60}$ fullerenes shell. As to the Eu@C$_{60}$, Giant autoionization leads to fast photoelectrons (about hundred eV) that go out almost untouched by the C$_{60}$ shell.

As a result of the outgoing electrons energy difference the atomic Giant resonances will be largely destroyed in A@C$_{60}$ while the Giant autoionization resonance will be almost completely preserved. Thus, on the way from Xe@C$_{60}$ Giant resonance to Eu@C$_{60}$ Giant autoionization resonance the oscillation structure should disappear. Similar will be the decrease of oscillations on the way from pure Giant to pure Giant autoionization resonances for the angular anisotropy parameters.

At Giant resonance frequencies the role of polarization of the fullerenes shell by the incoming photon beam is inessential.

Quite different is the situation for the outer electrons in Eu@C$_{60}$, the photoionization of which will be also considered.

1. **Introduction**

It was demonstrated recently that fullerenes electron shells strongly modify the Giant atomic resonances. As an example, the photoionization cross section of $4d$ electrons of Xe atom, "caged" inside the C$_{60}$ fullerene (Xe@C$_{60}$) was considered. It has been demonstrated that instead of profound $4d$ Giant resonance (GR) in Xe, it transforms into a sequence of several maxima in the endohedral atom Xe@C$_{60}$ [1].

This modification is a clear manifestation of the reflection and refraction of the relatively slow photoelectrons from $4d$ by the fullerenes shell in Xe. It is essential that the atomic GRs are located at so high frequencies that they are almost not affected by modification of the incoming photon beam by the polarized by this same beam C$_{60}$ electron shell.

All atoms of the Periodic table from I, Xe up to Eu have big powerful maxima in their photoabsorption cross-sections. However, it is known since long ago that these maxima on the way from I and Xe to Eu essentially modify their microscopic nature [2]. Indeed, for I and Xe this maxima are Giant resonances that are manifestations of correlations between all ten $4d$ electrons[1]. The photoelectrons formed by the Giant resonance are from the same $4d$ subshell and therefore have relatively low energy of about two Rydbergs.

In Eu with its semi-filled $4f$ subshell the photoabsorption cross section is instead a pure Giant *autoionization* resonance that originates from decay of a very powerful discrete transition of $4d$ electron to an empty $4f$ level, $4d \rightarrow 4f$, into the continuous spectrum excitation of the $4f$ electrons, $4f \rightarrow \varepsilon g, d$. As a result, the photoelectrons originate not from $4d$ as in Xe but from the outer $4f$ subshell. Therefore, they have much higher energy, of about ten Rydbergs and other angular momenta than photoelectrons from $4d$ in Xe. On the way from Xe to Eu the role of pure Giant resonance decreases while the contribution of autoionization decay grows up. There are indications that already in Ce the autoionization portion is dominant.

This difference in photoelectron energies has profound effect upon the photoabsorption cross sections of endohedral atoms near $4d$ threshold, since slow electrons are reflected by the fullerenes shell while the fast are not. As a result, one should expect strong modification of the Giant $4d$ resonance, while $4d$ Giant *autoionization* resonance remains untouched.

When inside the fullerene, the "caged" atom of the Lanthanides group can lose one or two of its electrons to the fullerenes shell. As a result, the oscillator strength of the $4d \rightarrow 4f$ transition increases, thus leading to Giant autoionization instead of a Giant resonance with corresponding increase of the photoelectron energy. As it was already mentioned, the increase of photoionization energy leads to elimination of the oscillations in the photoabsorption of endohedral atoms as compared to the isolated ones.

To check this conception, we have performed calculations of the partial and total cross-section of photoionization of Eu@$C_{60}$ and compared this with cross-sections for atomic Eu [3] and Xe@$C_{60}$. The research presented in this paper was stimulated by the publication of R. Phaneuf and his group [4][2], where they measured the photoabsorption cross-section of Ce@$C_{82}^+$ in order to find variations in the $4d$ region similar to that predicted in [1], and failed. In their analyses they assumed that inside the fullerenes cage Ce is stripped off its three outer electrons.

We believe now that the result obtained in this paper explains the absence of visible fine structure in the photoabsorption cross-section in the endohedral Ce@$C_{82}^+$ in the Ce Giant resonance region. The decisive answer can, however, come from electron spectroscopy research with coincidental measurement of the photoelectron energy and the ion yield.

We present here also the results for the outer shells of Eu@$C_{60}$, where similar to the case of Xe@$C_{60}$ new so-called Giant endohedral resonances are formed [6] due to combined action of two factors: the reflection of the photoelectron from the "caged" atom and the amplification of the electromagnetic field acting upon this atom. The new Giant endohedral resonances in the outer shells appear stronger or weaker in all endohedrals, so that we call the phenomenon Giant resonance resurrection.

All this discussion is of great general interest and value, since Giant resonances are universal features of the excitation of any finite many-fermion systems: nuclei, atoms, fullerenes, and clusters. They represent collective, coherent oscillations of many particles and manifest themselves most prominently in photon absorption cross sections. In a nucleus Giant resonances represent the excitation of coherent oscillatory motion of all protons relative to all neutrons [7], while in all other objects mentioned above they represent the coherent motion of all electrons of at least one many-electron shell (in atoms) and all collective electrons in metallic clusters and fullerenes relative to the atomic nucleus or the positive charge of a number of nuclei. Giant resonances are manifestations of plasmon-type or Langmuir excitations in a homogeneous electron gas [8] or so-called "zero" sound in a Fermi-liquid [9].

---

[1] Often erroneously it is called "shape resonance". This name is misleading since ignores the real multi-electron origin of the observed maxima.
[2] Some data on Ce@$C_{82}$ can be found in [5].

The universal nature of the Giant resonances (GRs) was recently emphasized in [10]. In photoabsorption cross sections as functions of photon energy, GRs, are represented by huge broad maxima. They have very large so-called oscillator strengths, which are determined by the total integrated area of the photoabsorption cross-section curves. The similarity of GRs in different objects is amazing, since when plotted on the same relative scales, GRs in nuclear Pb, $4d^{10}$ subshell in atomic Xe and in fullerene $C_{60}$ look alike [10].

It is important to emphasize that the ratio of the resonance width $\Gamma$ that characterizes its lifetime $\tau$, $\tau \sim 1/\Gamma$, to the frequency $\Omega$ is almost the same for all the above-mentioned objects, $1/5 \leq \Gamma/\Omega \leq 1/4$. It is obvious that the absolute values of the resonance energies and cross-sections differ in these objects by orders of magnitude, particularly when we compare the values for the $4d^{10}$ subshell in atomic Xe and nuclear Pb.

Therefore, it is of special interest and quite instructive to find a system, in which relatively small variation – substitution of one "caged" atom by another affect the cross-section so prominently. It is of interest to note that the possible difference in the Giant resonance structure in Xe@$C_{60}$ and Eu@$C_{60}$ was noted already in the introduction to our paper [1], where it was written: "The effect of $C_{60}$ rapidly disappears with the growth of the photoelectron energy $\varepsilon$. This is why Giant Autoionization resonances, e.g. those in Eu [3], remain unaffected by the $C_{60}$ shell, since relatively fast photoelectrons are emitted after their excitation".

## 2. Main formulas

We will use here the theoretical approaches already developed in a number of previous papers [11-13]. However, for completeness, let us repeat the main points of the consideration and present the essential formula used in calculations.

Let us start with the problem of an isolated closed shell atom. The following relation gives the differential in angle photoionization cross-section of an atom by non-polarized light of frequency $\omega$ in the dipole approximation (see e.g. [14]):

$$\frac{d\sigma_{nl}(\omega)}{d\Omega} = \frac{\sigma_{nl}(\omega)}{4\pi}[1 - \frac{\beta_{nl}}{2}P_2(\cos\theta)], \qquad (1)$$

where $\kappa = \omega/c$, $P_2(\cos\theta)$ is the second order Legendre polynomial, $\theta$ is the angle between photon momentum $\kappa$ and photoelectron velocity $\mathbf{v}$, $\beta_{nl}(\omega)$ is the dipole angular anisotropy parameter.

There are two possible dipole transitions from subshell $l$, namely $l \to l \pm 1$. In one-electron Hartree-Fock (HF) approximation the cross-section $\sigma_{nl}(\omega)$ is given by the expression

$$\sigma_{nl}(\omega) = \frac{4\pi^2}{\omega c}(2l+1)[(l+1)d^2_{l+1} + ld^2_{l-1}] \equiv \sigma_{nl,\varepsilon l+1}(\omega) + \sigma_{nl,\varepsilon l-1}(\omega), \qquad (2)$$

while $\beta_{nl}(\omega)$ parameter is presented as [15]:

$$\beta_{nl}(\omega) = \frac{1}{(2l+1)\left[(l+1)d_{l+1}^2 + ld_{l-1}^2\right]}[(l+1)(l+2)d_{l+1}^2 + l(l-1)d_{l-1}^2 -$$
$$6l(l+1)d_{l+1}d_{l-1}\cos(\delta_{l+1} - \delta_{l-1})]. \qquad (3)$$

Here $\delta_l(k)$ are the photoelectrons' scattering phases. The following relation gives the matrix elements $d_{l\pm 1}$ in the so-called *r*-form

$$d_{l\pm 1} \equiv \int_0^\infty P_{nl}(r) r P_{\varepsilon l \pm 1}(r) dr, \qquad (4)$$

where $P_{nl}(r)$, $P_{\varepsilon l\pm 1}(r)$ are the radial Hartree-Fock (HF) [14] one-electron wave functions of the *nl* discrete level and $\varepsilon l \pm 1$ - in continuous spectrum, respectively.

In order to take into account the Random Phase Approximation with Exchange (RPAE) [14] multi-electron correlations, one has to perform the following substitutions in the expression for $\beta_{nl}(\omega)$ [15]:

$$d_{l+1}d_{l-1}\cos(\delta_{l+1} - \delta_{l-1}) \to [(\operatorname{Re} D_{l+1} \operatorname{Re} D_{l-1} + \operatorname{Im} D_{l+1} \operatorname{Im} D_{l-1})\cos(\delta_{l+1} - \delta_{l-1}) -$$
$$- (\operatorname{Re} D_{l+1} \operatorname{Im} D_{l-1} - \operatorname{Im} D_{l+1} \operatorname{Re} D_{l-1})\sin(\delta_{l+1} - \delta_{l-1})] \equiv$$
$$\equiv \tilde{D}_{l+1}\tilde{D}_{l-1}\cos(\delta_{l+1} + \Delta_{l+1} - \delta_{l-1} - \Delta_{l-1}). \qquad (5)$$
$$d_{l\pm 1}^2 \to \operatorname{Re} D_{l\pm 1}^2 + \operatorname{Im} D_{l\pm 1}^2 \equiv \tilde{D}_{l\pm 1}^2$$

Here the following notations are used for the matrix elements with account of multi-electron correlations

$$D_{l\pm 1}(\omega) \equiv \tilde{D}_{l\pm 1}(\omega)\exp[i\Delta_{l\pm 1}(\varepsilon)], \qquad (6)$$

where $\tilde{D}_{l\pm 1}(\omega)$ and $\Delta_{l\pm 1}$ are absolute values of the amplitudes for respective transitions and phases for photoelectrons with angular moments $l\pm 1$.

The following are the RPAE equation for the dipole matrix elements [14]

$$\langle \nu_2 | D(\omega) | \nu_1 \rangle = \langle \nu_2 | d | \nu_1 \rangle + \sum_{\nu_3,\nu_4} \frac{\langle \nu_3 | D(\omega) | \nu_4 \rangle (n_{\nu_4} - n_{\nu_3}) \langle \nu_4 \nu_2 | U | \nu_3 \nu_1 \rangle}{\varepsilon_{\nu_4} - \varepsilon_{\nu_3} + \omega + i\eta(1 - 2n_{\nu_3})}, \qquad (7)$$

where

$$\langle \nu_1 \nu_2 | \hat{U} | \nu_1' \nu_2' \rangle \equiv \langle \nu_1 \nu_2 | \hat{V} | \nu_1' \nu_2' \rangle - \langle \nu_1 \nu_2 | \hat{V} | \nu_2' \nu_1' \rangle. \qquad (8)$$

Here $\hat{V} \equiv 1/|\vec{r} - \vec{r}'|$ and $\nu_i$ is the total set of quantum numbers that characterize a HF one-electron state on discrete (continuum) levels. That includes the principal quantum number (energy), angular momentum, its projection and the projection of the electron spin. The function $n_{v_i}$ (the so-called step-function) is equal to 1 for occupied and 0 for vacant states.

The dipole matrix elements $D_{l\pm 1}$ are obtained by solving the radial part of the RPAE equation (12) numerically, using the procedure discussed at length in [16].

In this paper we consider also half-filled shell atom Eu. This was never done before for endohedrals. However the generalization is straightforward. As it was discussed in a number of our papers, (see, e.g. [17]), we treat such an atom as having two types of electrons, namely "up" and "down", denoted with an arrow $\uparrow$ and $\downarrow$, respectively. Due to presence of the semi–filled level $4f^7 \uparrow$, each subshell splits into two levels, up" $\uparrow$ and "down" $\downarrow$ with different ionization potentials and without exchange between these electrons. As a result, each cross-section, angular anisotropy parameter, dipole matrix elements, one-electron wave functions and scattering phase of a given level become spin-dependent values $\sigma_{nl\uparrow,\downarrow}(\omega)$, $\beta_{nl\uparrow,\downarrow}(\omega)$, $D_{l\pm 1\uparrow,\downarrow}(\omega)$, $d_{l\pm 1\uparrow,\downarrow}$, $\delta_{l\pm 1\uparrow,\downarrow}$, and $\Delta_{l\pm 1\uparrow,\downarrow}$. While the photon interaction cannot connect the spin "up" and "down" states, the Coulomb interelectron interaction connects them. As a result, instead of integral linear equation (6), one has to solve a matrix integral equation that is symbolically presented in the following form:

$$\left(\hat{D}_\uparrow(\omega)\hat{D}_\downarrow(\omega)\right) = \left(\hat{d}_\uparrow(\omega)\hat{d}_\downarrow(\omega)\right) + \left(\hat{D}_\uparrow(\omega)\hat{D}_\downarrow(\omega)\right) \times \begin{pmatrix} \hat{\chi}_{\uparrow\uparrow}(\omega) & 0 \\ 0 & \hat{\chi}_{\downarrow\downarrow}(\omega) \end{pmatrix} \times \begin{pmatrix} \hat{U}_{\uparrow\uparrow} & \hat{V}_{\uparrow\downarrow} \\ \hat{V}_{\downarrow\uparrow} & \hat{U}_{\downarrow\downarrow} \end{pmatrix} \qquad (9)$$

In these same notations (7) is presented as

$$\hat{D}(\omega) = \hat{d} + \hat{D}(\omega) \times \hat{\chi}(\omega) \times \hat{U} \qquad (10)$$

## 3. Effect of C$_{60}$ fullerene shell

Let us start with the effects of photoelectron reflection. These effects near the photoionization threshold can be described within the framework of the "orange" skin potential model. According to this model, for small photoelectron energies the real static potential of the C$_{60}$ can be presented by the zero-thickness bubble pseudo-potential (see [18, 19] and references therein):

$$V(r) = -V_0 \delta(r - R). \qquad (11)$$

The parameter $V_0$ is determined by the requirement that the binding energy of the extra electron in the negative ion $C_{60}^-$ is equal to its observable value. Addition of the potential (11) to the atomic HF potential leads to a factor $F_l(k)$ in the photoionization amplitudes, which depends only upon the photoelectron's momentum $k$ and orbital quantum number $l$ [18, 19]:

$$F_l(k) = \cos \breve{\Delta}_l(k) \left[ 1 - \tan \breve{\Delta}_l(k) \frac{v_{kl}(R)}{u_{kl}(R)} \right], \tag{12}$$

where $\breve{\Delta}_l(k)$ are the additional phase shifts due to the fullerene shell potential (11). They are expressed by the following formula:

$$\tan \breve{\Delta}_l(k) = \frac{u_{kl}^2(R)}{u_{kl}(R)v_{kl}(R) + k/2V_0}. \tag{13}$$

In these formulas $u_{kl}(r)$ and $v_{kl}(r)$ are the regular and irregular solutions of the atomic HF equations for a photoelectron with momentum $k = \sqrt{2\varepsilon}$, where $\varepsilon$ is the photoelectron energy connected with the photon energy $\omega$ by the relation $\varepsilon = \omega - I_A$ with $I$ being the atom A ionization potential.

Using Eq. (12), one can obtain the following relation for $D^{AC(r)}$ and $Q^{AC(r)}$ amplitudes for endohedral atom A@C$_{60}$ with account of photoelectron's reflection and refraction by the C$_{60}$ static potential (11), expressed via the respective values for isolated atom that correspond to $nl \to \varepsilon l'$ transitions:

$$D_{nl,kl'}^{AC(r)}(\omega) = F_{l'}(k) D_{nl,kl'}(\omega), . \tag{14}$$

For the cross-sections one has

$$\sigma_{nl,kl'}^{AC(r)}(\omega) = [F_{l'}(k)]^2 \sigma_{nl,kl'}(\omega), \tag{15}$$

where $\sigma_{nl,kl'}(\omega)$ is the contribution of the $nl \to \varepsilon l'$ transition to the photoionization cross-section of atomic subshell $nl$, $\sigma_{nl}(\omega)$ [see (2)].

Now let us discuss the role of polarization of the C$_{60}$ shell under the action of the photon beam [20]. The effect of the fullerene electron shell polarization upon atomic photoionization amplitude can be taken into account in RPAE using (7). An essential simplification comes from the fact, that the C$_{60}$ radius $R_C$ is much bigger than the atomic radius $r_a$, $R_C \gg r_a$. It is also important that the electrons in C$_{60}$ are located within a layer, the thickness of which $\Delta_C$ is considerably smaller than $R_C$. In this case, the amplitude of endohedral atom's photoionization due to $nl \to \varepsilon l'$ transition with all essential atomic correlations taken into account can be presented by the following formula [20]:

$$D_{nl,\varepsilon l'}^{AC}(\omega) \cong F_{l'}(k) \left( 1 - \frac{\alpha_C^d(\omega)}{R_C^3} \right) D_{nl,\varepsilon l}(\omega) \equiv F_{l'}(k) G^d(\omega) D_{nl,\varepsilon l}(\omega), \tag{16}$$

where $\alpha_C^d(\omega)$ is the dipole dynamical polarizability of C$_{60}$ and $R_C$ is its fullerenes radius. The $G^d(\omega)$ factor is a complex number that we present as

$$G^d(\omega) = \tilde{G}^d(\omega)\exp[i\eta^d(\omega)], \qquad (17)$$

where $\tilde{G}^d(\omega)$ are respective absolute values.

Using the relation between the imaginary part of the polarizability and the dipole photoabsorption cross-section $\sigma_C^d(\omega)$ - $\operatorname{Im}\alpha_C^d(\omega) = c\sigma_C^d(\omega)/4\pi\omega$, one can derive the polarizability of the C$_{60}$ shell. Although experiments [21, 22] do not provide absolute values of $\sigma_C^d(\omega)$, it can be reliably estimated using different normalization procedures on the basis of the sun rule: $(c/2\pi^2)\int_{I_o}^{\infty}\sigma_C^d(\omega)d\omega = N$, where $N$ is the number of collectivized electrons. The real part of polarizability is connected with imaginary one (and with the photoabsorption cross-section $\sigma_C^d(\omega)$) by the dispersion relation:

$$\operatorname{Re}\alpha_C^d(\omega) = \frac{c}{2\pi^2}\int_{I_C}^{\infty}\frac{\sigma_C^d(\omega')d\omega'}{\omega'^2 - \omega^2}, \qquad (18)$$

where $I_C$ is the C$_{60}$ ionization potential. This approach was used for polarizability of C$_{60}$ in [23], where it was considered that $N = 240$, i.e. 4 collectivized electrons per each C atom in C$_{60}$. Using the photoabsorption data that are considered as most reliable in [21], we obtained $N_{eff} \approx 250$ that is sufficiently close to the value, assumed in [23].

Note that because of the strong inequality $R_C \gg r_a$ we have derived the formula (16) that is more accurate than that obtained from the RPAE for the whole endohedral atom system. This is important since "one electron – one vacancy" channel that is the only taken into account in RPAE is not always dominant in the photoabsorption cross-section of the fullerene and hence in its polarizability.

Using the definition (2), the amplitude (16) and performing the substitution (5), one has for the cross section

$$\sigma_{nl,\varepsilon l'}^{AC}(\omega) = [F_{l'}(\omega)]^2\left|1 - \frac{\alpha_C^d(\omega)}{R_C^3}\right|^2 \sigma_{nl,\varepsilon l'}(\omega) \equiv [F_{l'}(\omega)]^2 S(\omega)\sigma_{nl,\varepsilon l'}(\omega), \qquad (19)$$

where $S(\omega) = [\tilde{G}^d(\omega)]^2$ cab be called radiation enhancement parameter.

With these same amplitudes (16), using the expression (3) and performing the substitution (5) we obtain the angular anisotropy parameter for considered endohedrals. While calculating the anisotropy parameter, the cosine of atomic phase differences $\cos(\delta_l - \delta_{l'})$ in formula (3) is replaced by $\cos(\delta_l + \Delta_l - \delta_{l'} - \Delta_{l'})$. As a result, one has for the dipole angular anisotropy parameter:

$$\beta_{nl}(\omega) = \frac{1}{(2l+1)\left[(l+1)F_{l+1}^2\tilde{D}_{l+1}^2 + lF_{l-1}^2\tilde{D}_{l-1}^2\right]}[(l+1)(l+2)F_{l+1}^2\tilde{D}_{l+1}^2$$
$$+ l(l-1)F_{l-1}^2\tilde{D}_{l-1}^2 - 6l(l+1)F_{l+1}F_{l-1}\tilde{D}_{l+1}\tilde{D}_{l-1}\cos(\tilde{\delta}_{l+1} - \tilde{\delta}_{l-1})] \quad . \tag{20}$$

where $\tilde{\delta}_{l'} = \delta_{l'} + \Delta_{l'}$ (see (5)). Naturally, the dipole parameter $\beta_{nl}(\omega)$ is not affected by $G^d(\omega)$ factors that similarly alter the nominator and denominator in (20).

The radiation enhancement parameter $S(\omega) = [\tilde{G}^d(\omega)]^2$ is a characteristic of the fullerenes shell itself and is therefore the same for any atom "caged" inside the fullerene. The reflection factors $F_{l\pm1}(k)$ depends upon spin projection only indirectly, since for a given subshell $nl$ the $k$ values for a given $\omega$ are different since different are the ionization potentials $I_{nl\uparrow}$ and $I_{nl\downarrow}$.

## 4. Some details of calculations and their results

Naturally, the $C_{60}$ parameters in the present calculations were chosen the same as in the previous papers, e.g. in [24]: $R = 6.639$ and $V_0 = 0.443$.

We have calculated the cross-sections 5$p$, 5$s$, 4$d$ "up" and "down" electrons and 4$f$ "up" electrons in Eu@$C_{60}$ and compared the results with that for 5$p$, 5$s$, 4$d$ electrons in Xe@$C_{60}$. We have calculated also the dipole angular anisotropy parameter for 5$p$, 4$d$ "up" and "down" electrons and 4$f$ "up" electrons in Eu@$C_{60}$

We start by presenting in Fig. 1 the results for the radiation enhancement parameter $S(\omega)$. We have marked by arrows on the figure all thresholds of the considered subshells. It is seen that $S(\omega)$ rapidly approaches $S(\omega) = 1$ with photon frequency growth. To calculate the cross-sections, we need also the photoelectrons reflection parameters $F_{l\pm1}(\omega)$. In Fig. 2-5 we present these parameters for 5$p$, 5$s$, 4$d$ "up" and "down" electrons and 4$f$ "up". All curves are rapidly oscillating functions that within ten Ry from corresponding thresholds decreases the amplitude from approximately $0.5 \div 2$ and even bigger to $0.8 \div 1.2$ and even smaller. Fig.6 presents this parameter for 4$d$ in Xe, which is close to the corresponding value for Eu.

Figures 7 - 26 are dedicated to photoionization cross-sections. In Fig. 7, 8 we depict the results for 5$p$ "down" and "up" electrons in Eu and Eu@$C_{60}$, while Fig.9 gives the cross-section for 5$p$ electrons of Xe and Xe@$C_{60}$. We see that due to reflection and radiation enhancement parameters prominent maxima appear at about 2.8 Ry. The heights of maxima are of about 50 Mb that is big in atomic scale. They can be called Giant endohedral resonances [6, 25]. However, these maxima are much smaller than that in Xe [25]. Fig. 10 – 13 illustrates the effect of reflection factors upon partial photoionization cross-sections of 5$p$ "down" and "up" electrons in Eu@$C_{60}$ in comparison to the data in Eu.

Fig.14 and 15 presents the results for 5$s$ "down" and "up" electrons in Eu and Eu@$C_{60}$. It is seen, that above 5 Ry only reflection factor is essential and even its role above about 9 Ry is negligible.

Fig.16 depicts the cross-section of 4$f$ "up" electrons in Eu and Eu@$C_{60}$. At 2 Ry this cross – section has a Giant endohedral resonance that is a result of combined action of reflection and radiation enhancement. It has also a second maximum that is the Giant autoionization resonance, which is almost not affected by the fullerenes shell. It is essential to keep in mind that, as it was already said above, this is the main decay channel for the $4d\uparrow \to 4f\uparrow$ discrete transition. This cross-section is compared to the photoionization cross-section of 4d electrons in Xe and Xe@$C_{60}$ given by Fig. 17 that represent the atomic and destroyed by fullerenes shell Giant resonance.

Fig.18 and 19 depicts the cross-sections of 4f "up" electrons in Eu@$C_{60}$ and Eu atom, where only the effect of reflection is taken into account. Figures 20 – 25 present the data for 4d "down" and "up" electrons in Eu and Eu@$C_{60}$. At such high $\omega$ the role of radiation enhancement is negligible.

Fig.26 depicts the photoionization cross-section in the Giant autoionization resonance photon energy region of Eu and Eu@$C_{60}$. It is seen, as it was already suggested on the bases of qualitative arguments in the Introduction, that this resonance is almost unaffected by the fullerene shell. Most probable that this gives qualitative explanation of the lack of oscillations in the photoionization cross-section in the 4d frequency region that was observed in Ce@$C_{82}^+$ in [4] and for Ce@$C_{82}$ with considerably lower accuracy in [5].

Figures 27 – 27 present the angular anisotropy parameters of 5p, 4f, and 4d electrons for Eu and Eu@$C_{60}$. The fullerene shell affects them mainly close to the threshold. Modification in the energy range of the Giant autoionization resonance affects angular anisotropy parameter negligibly.

It would be of interest to see the alteration of the photoionization cross-section if instead of $C_{60}$ other fullerenes, like $C_{70}$, $C_{76}$, $C_{82}$ or $C_{87}$ are considered.

## Acknowledgements

This work was supported the Hebrew University Intramural Fund.

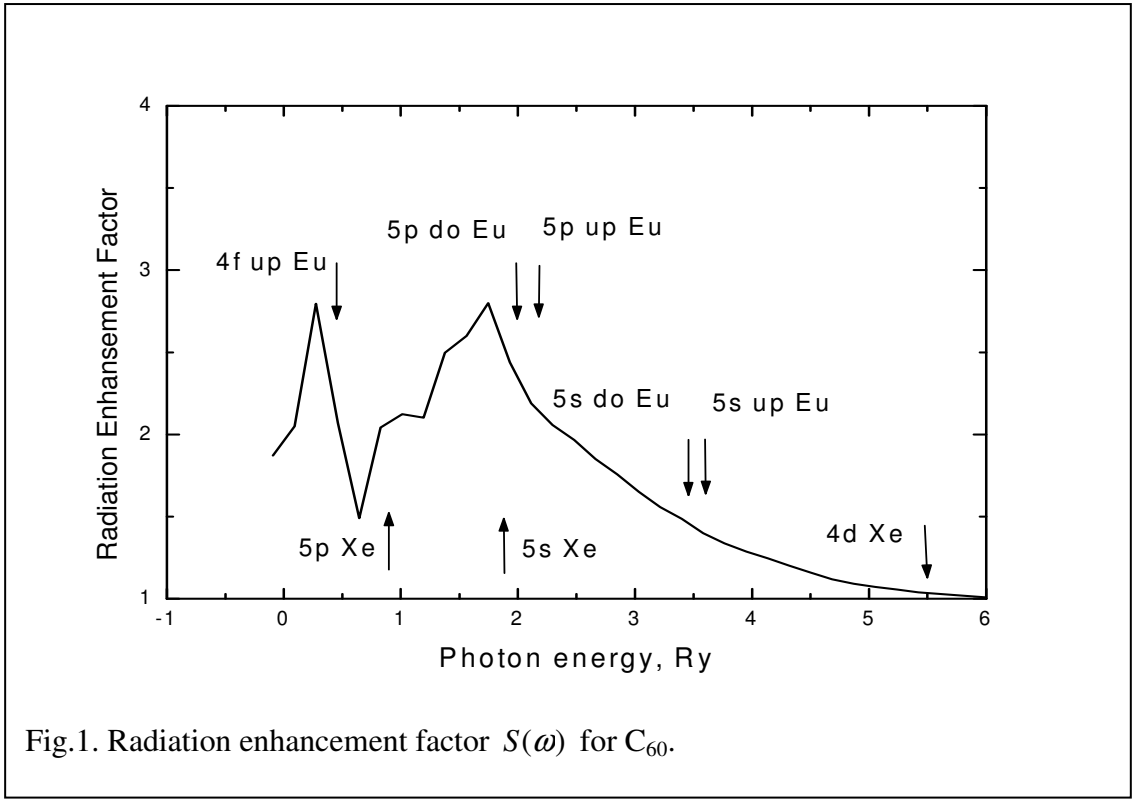

Fig.1. Radiation enhancement factor $S(\omega)$ for $C_{60}$.

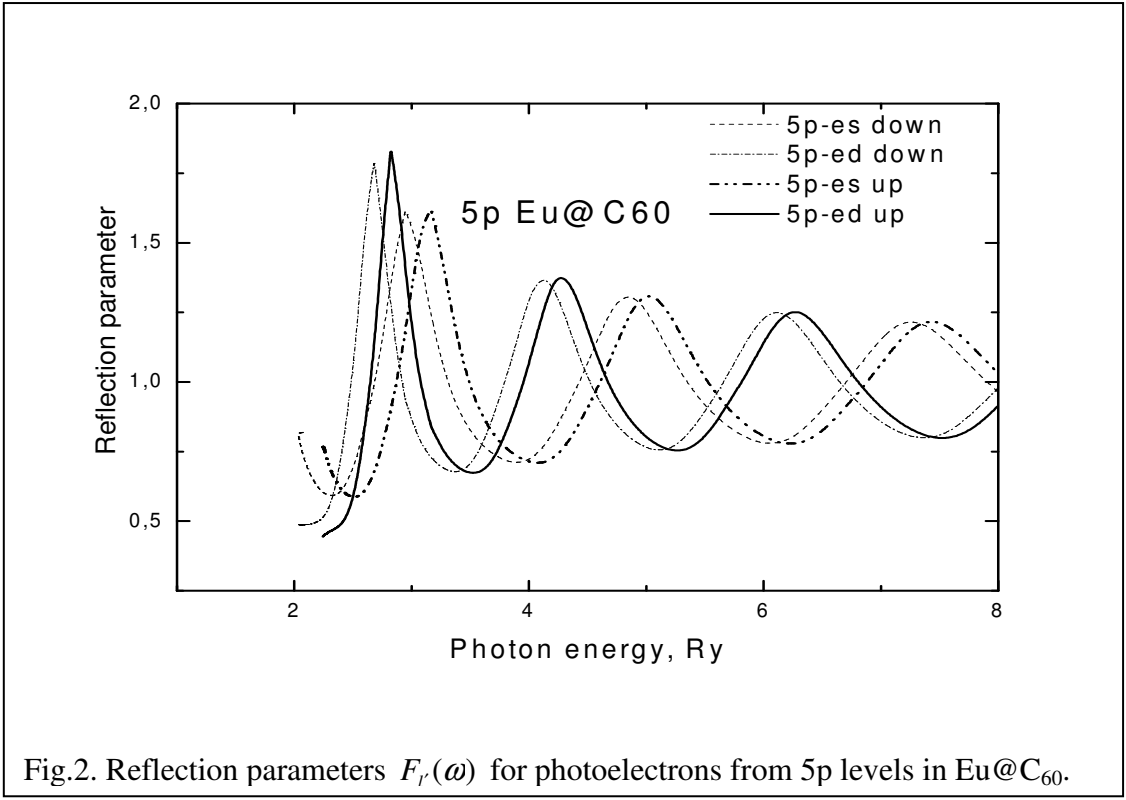

Fig.2. Reflection parameters $F_{l'}(\omega)$ for photoelectrons from 5p levels in Eu@$C_{60}$.

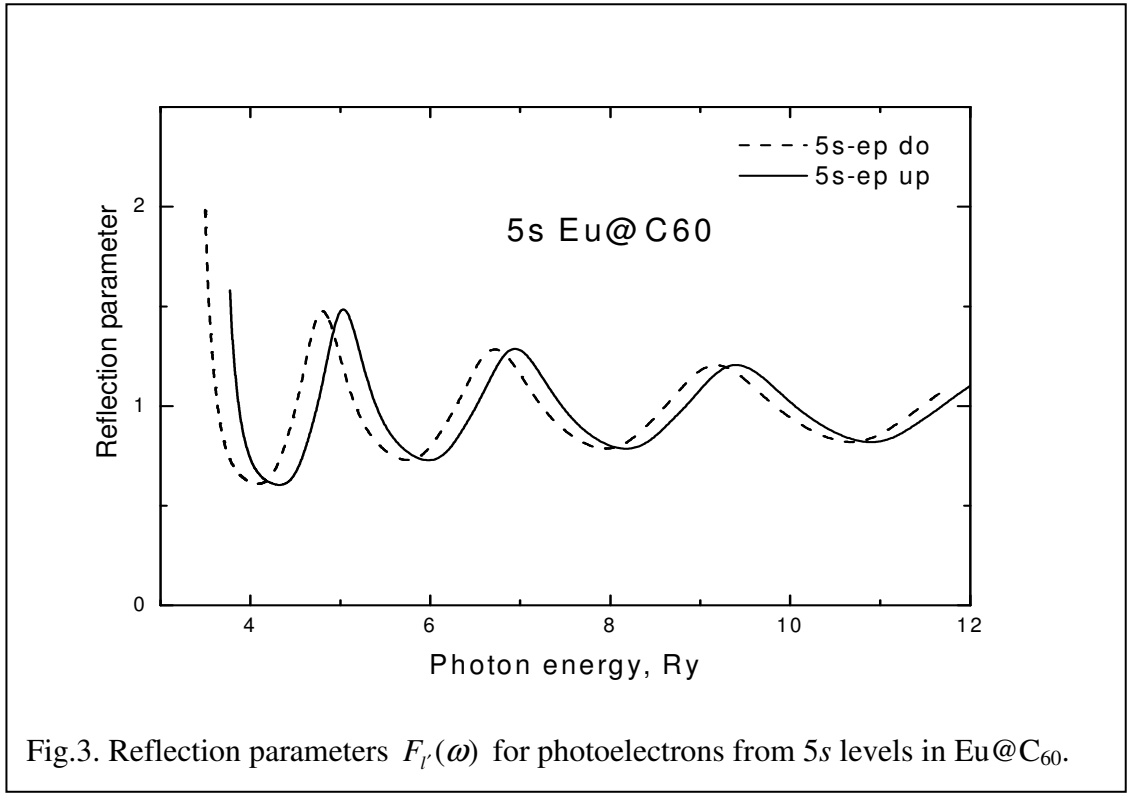

Fig.3. Reflection parameters $F_{l'}(\omega)$ for photoelectrons from 5$s$ levels in Eu@C$_{60}$.

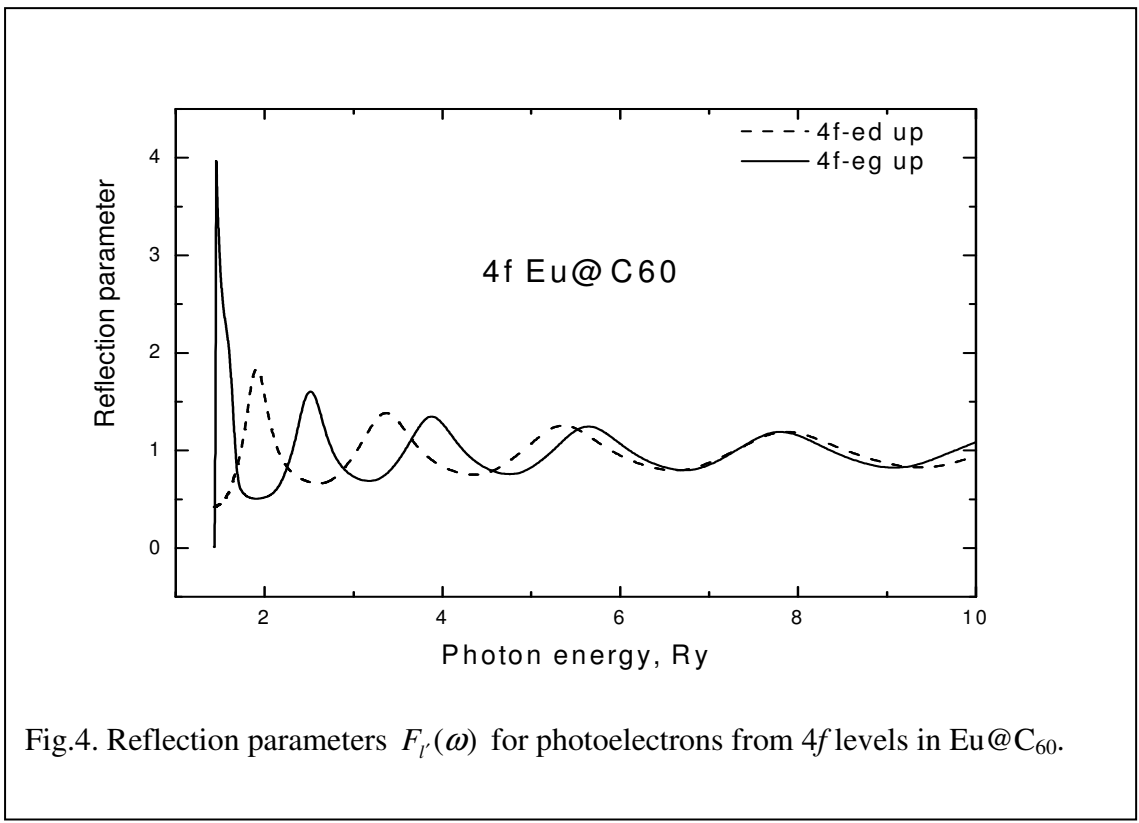

Fig.4. Reflection parameters $F_{l'}(\omega)$ for photoelectrons from 4$f$ levels in Eu@C$_{60}$.

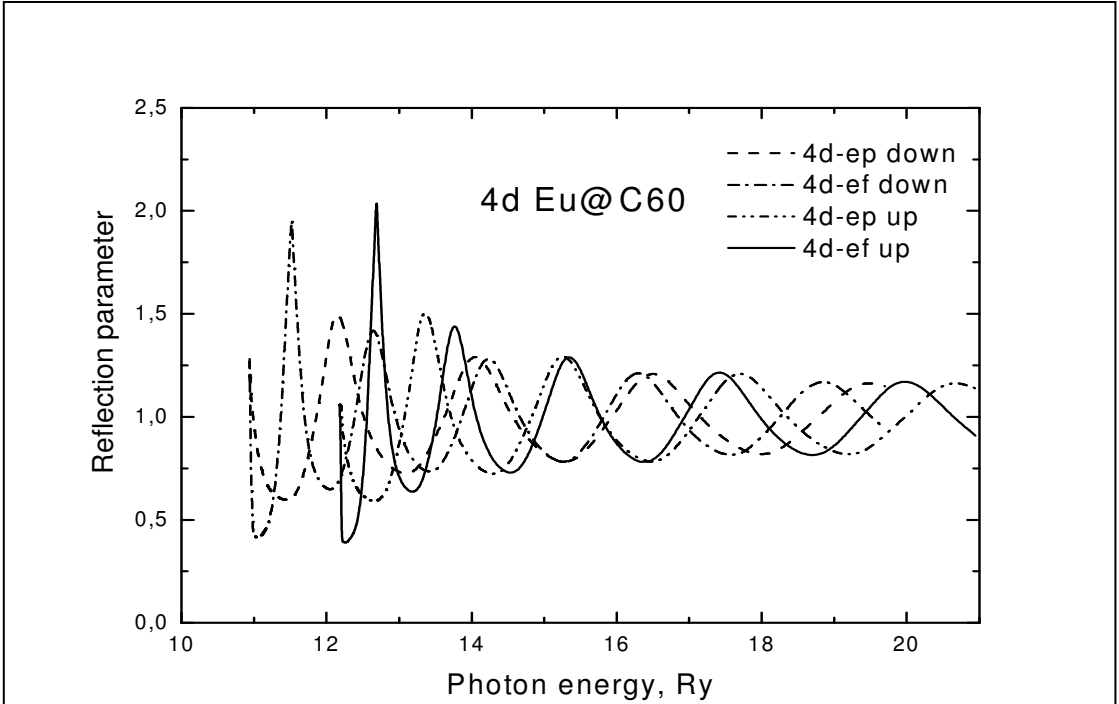

Fig. 5. Reflection parameters $F_{l'}(\omega)$ for photoelectrons from 4$d$ levels in Eu@C$_{60}$.

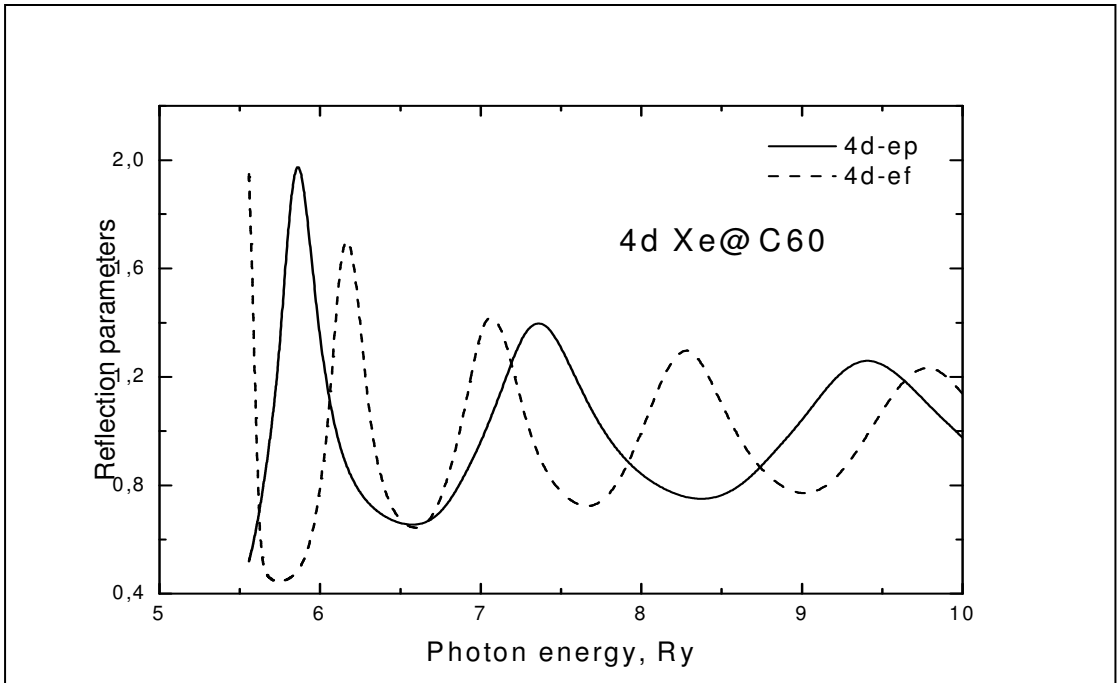

Fig.6. Reflection parameters $F_{l'}(\omega)$ for photoelectrons from 4$d$ levels in Xe@C$_{60}$.

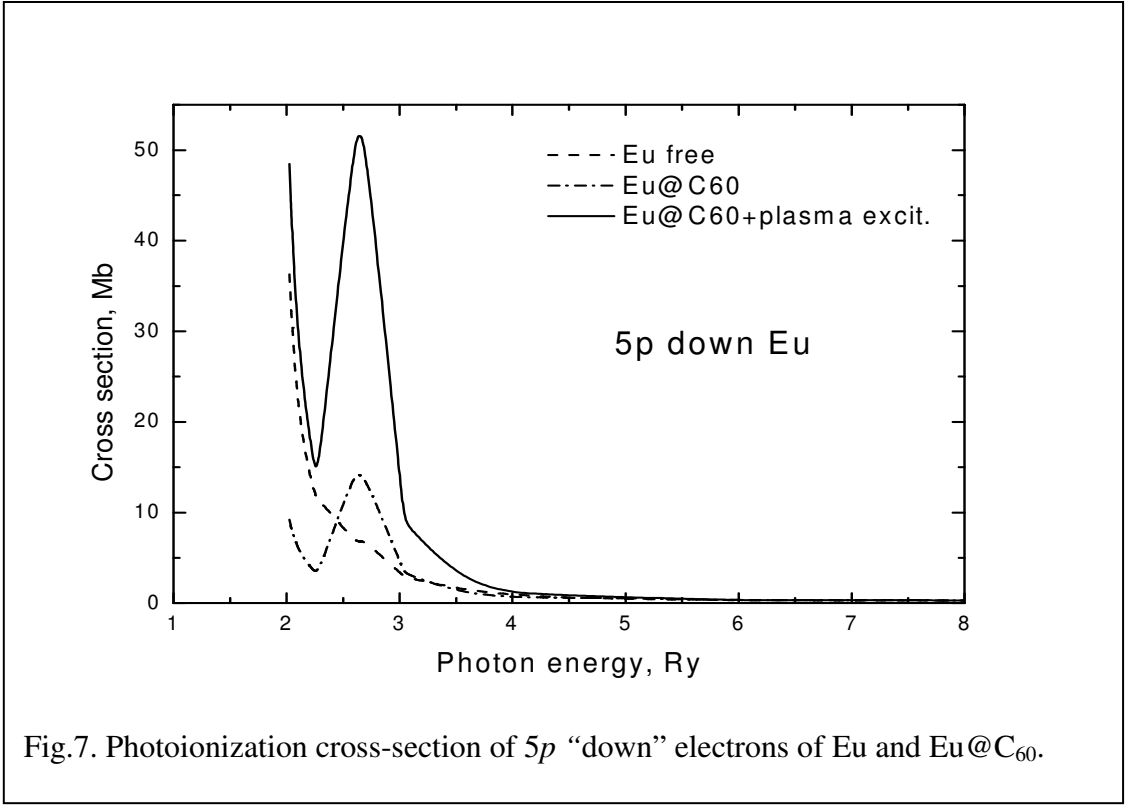

Fig.7. Photoionization cross-section of 5*p* "down" electrons of Eu and Eu@$C_{60}$.

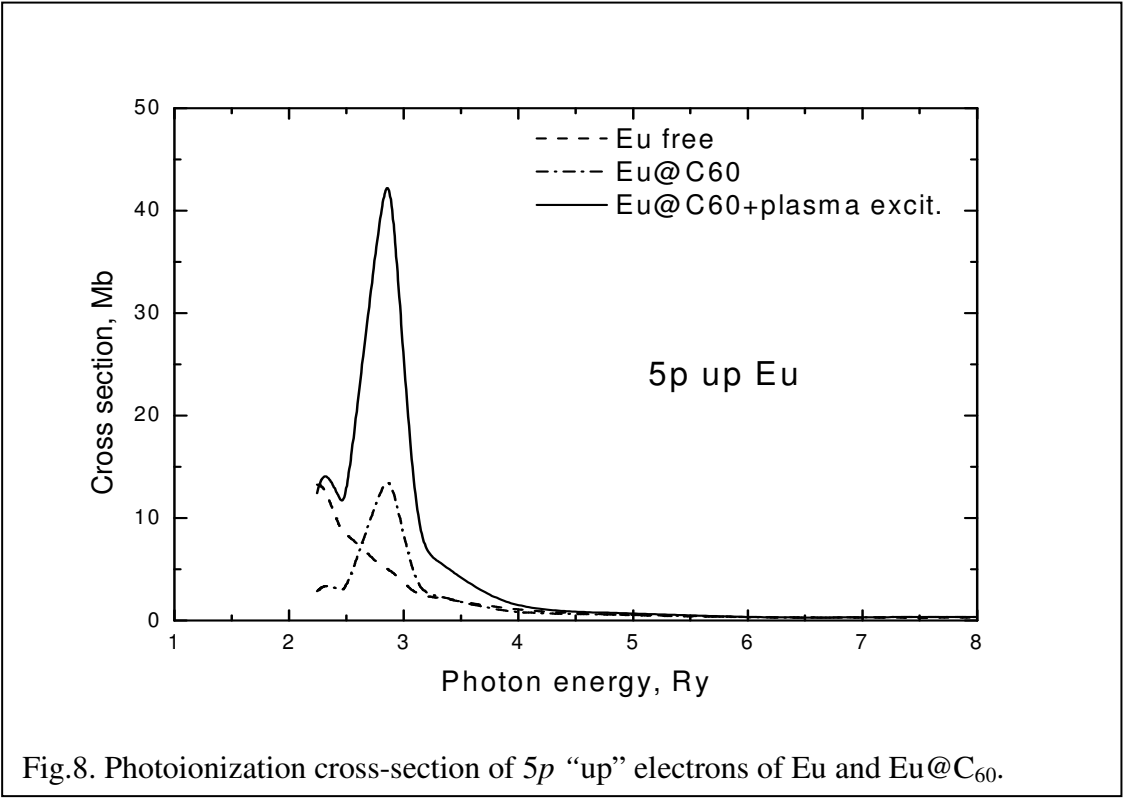

Fig.8. Photoionization cross-section of 5*p* "up" electrons of Eu and Eu@$C_{60}$.

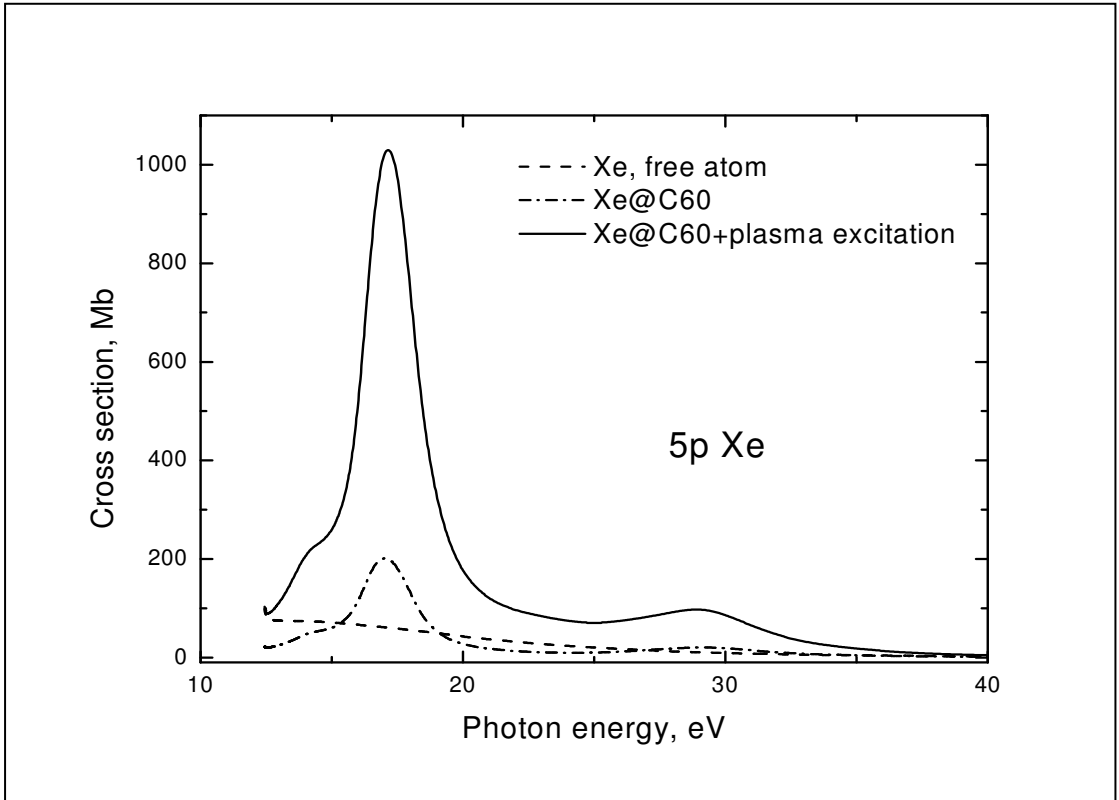

Fig.9. Photoionization cross-section of 5p electrons of Xe and Xe@$C_{60}$ [25].

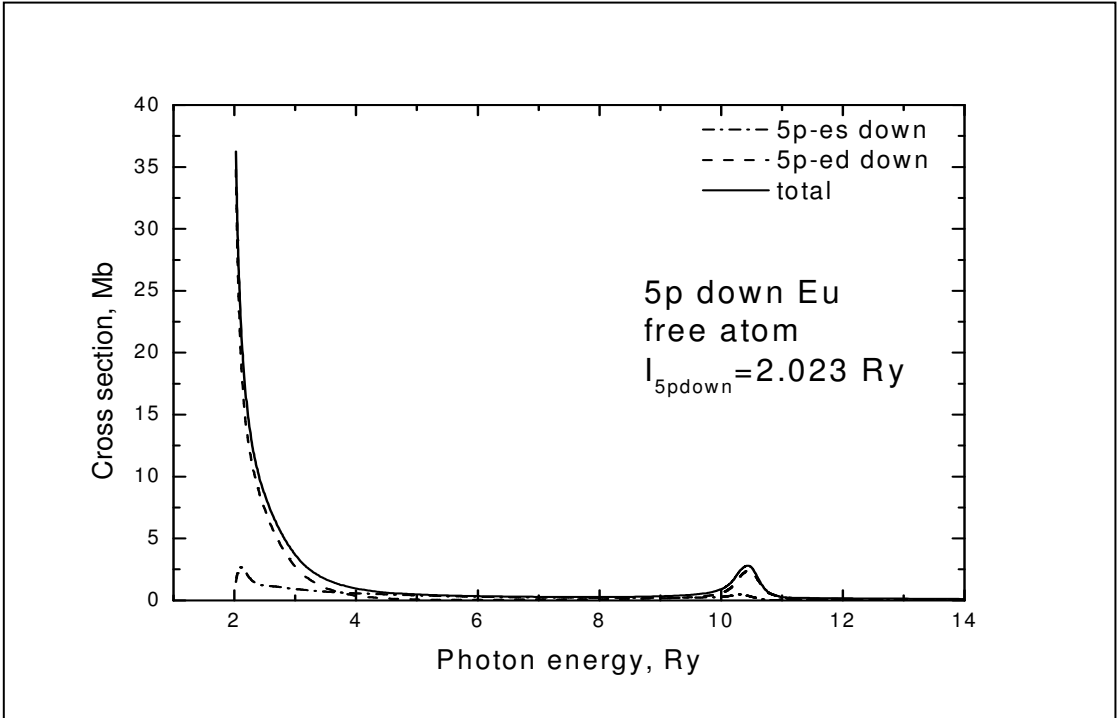

Fig.10. Photoionization cross-section of 5p "down" electrons of Eu atom.

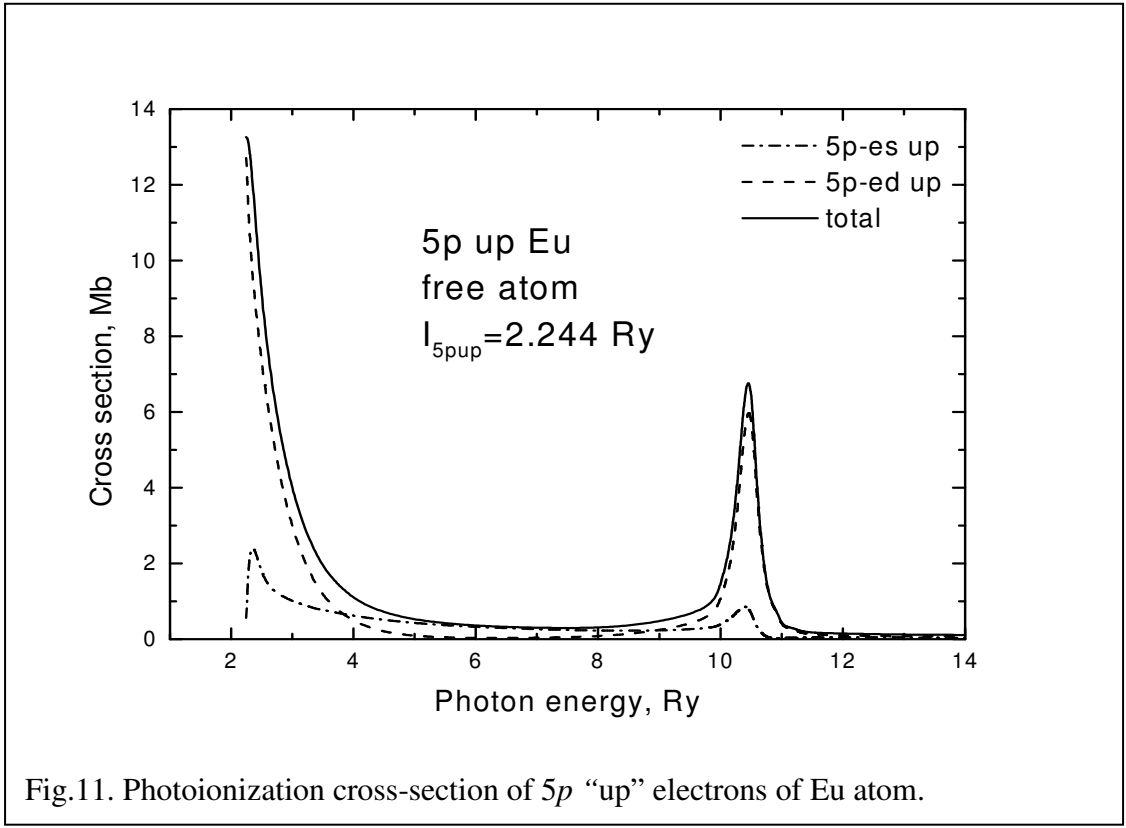

Fig.11. Photoionization cross-section of 5*p* "up" electrons of Eu atom.

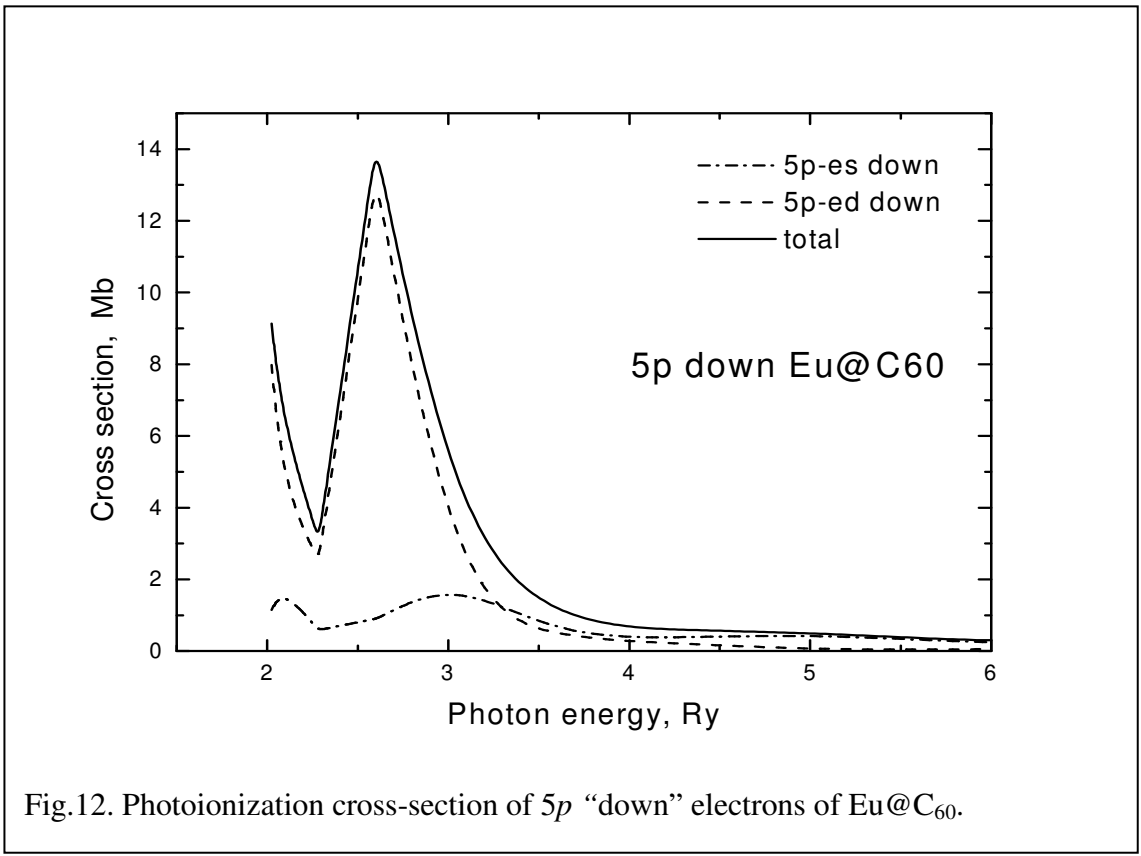

Fig.12. Photoionization cross-section of 5*p* "down" electrons of Eu@$C_{60}$.

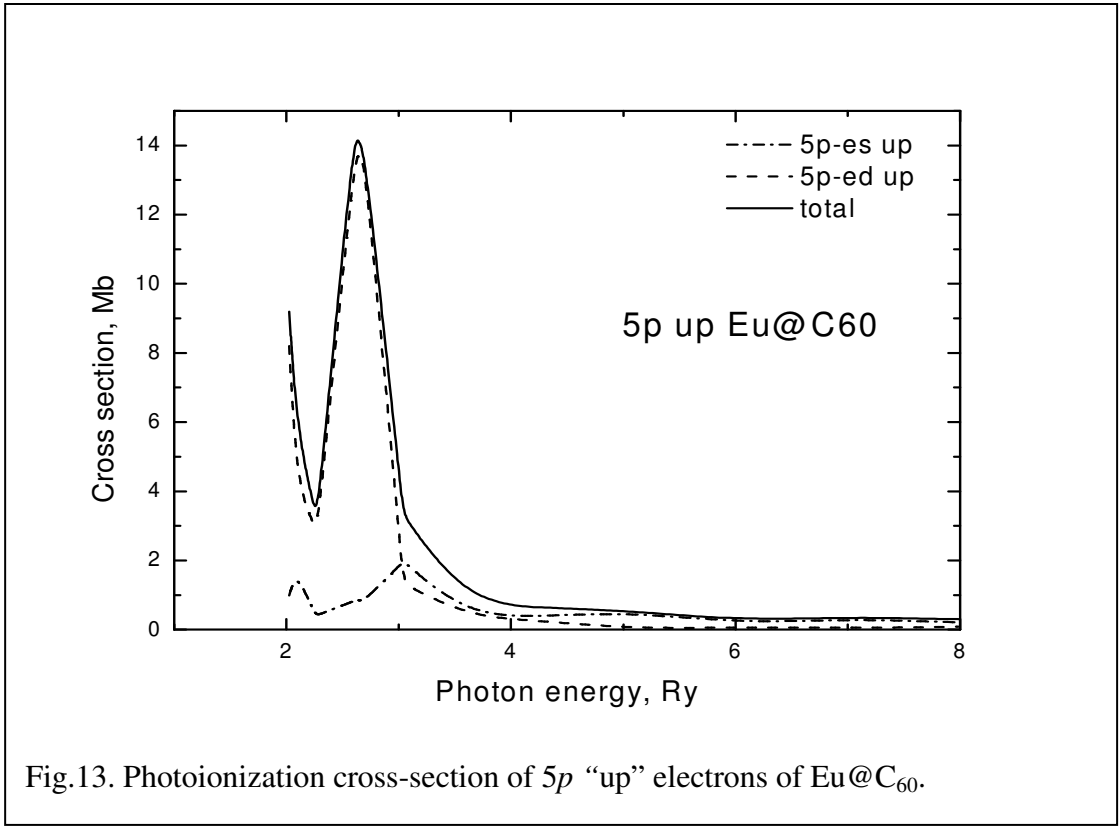

Fig.13. Photoionization cross-section of $5p$ "up" electrons of Eu@C$_{60}$.

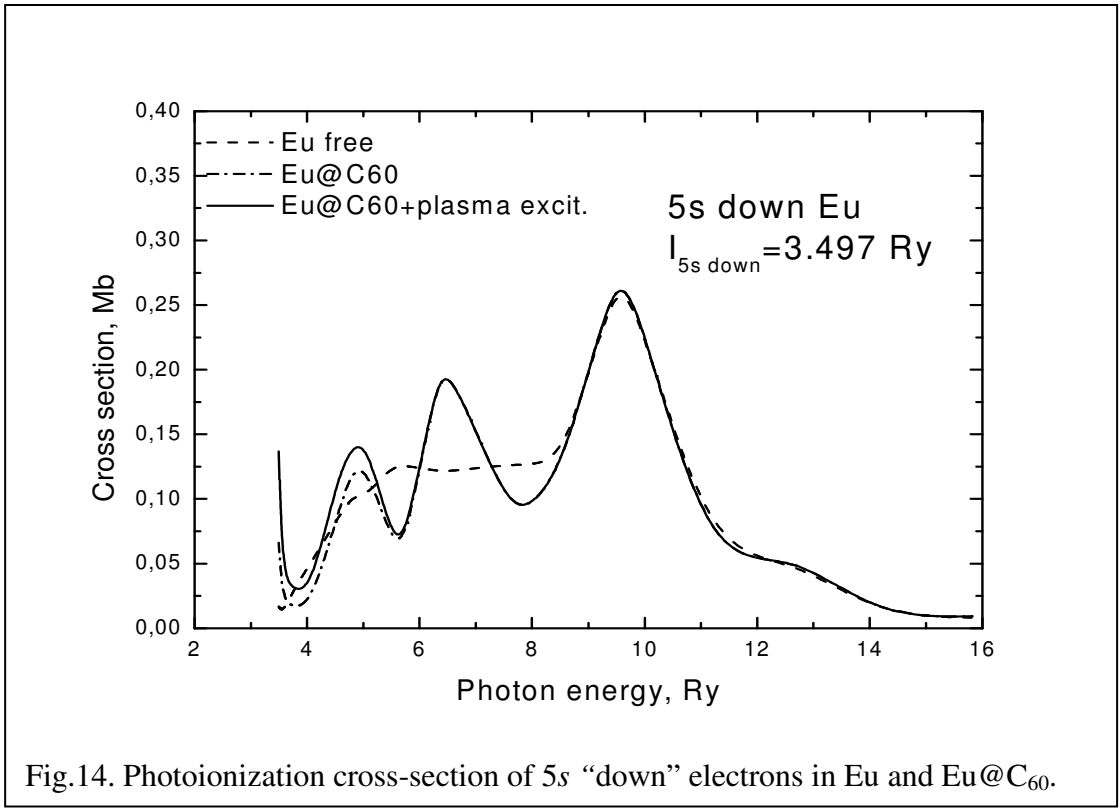

Fig.14. Photoionization cross-section of $5s$ "down" electrons in Eu and Eu@C$_{60}$.

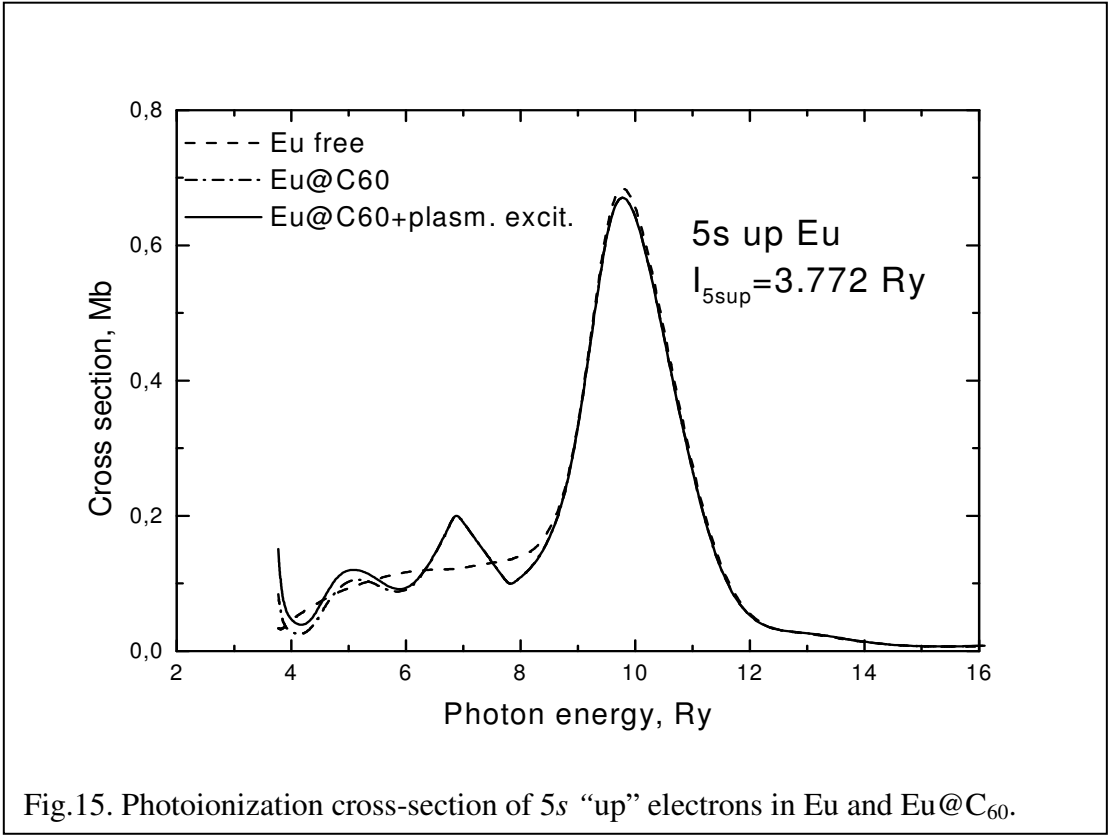

Fig.15. Photoionization cross-section of 5*s* "up" electrons in Eu and Eu@C$_{60}$.

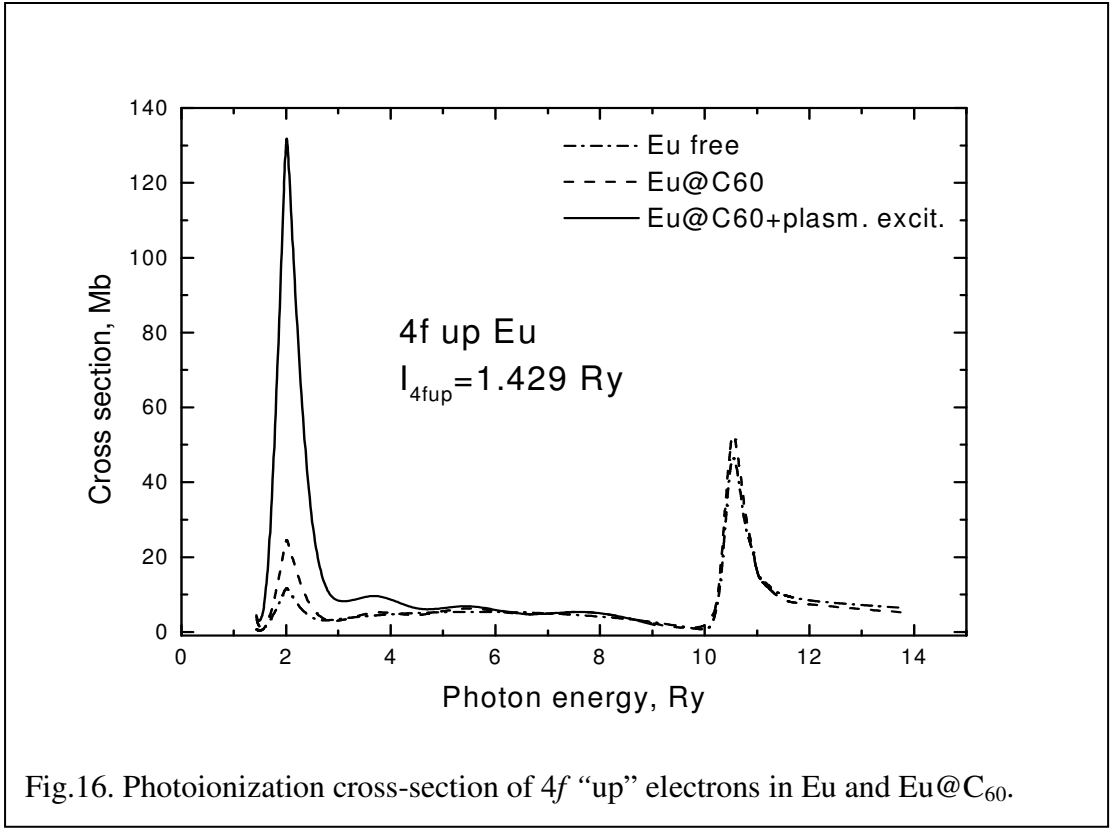

Fig.16. Photoionization cross-section of 4*f* "up" electrons in Eu and Eu@C$_{60}$.

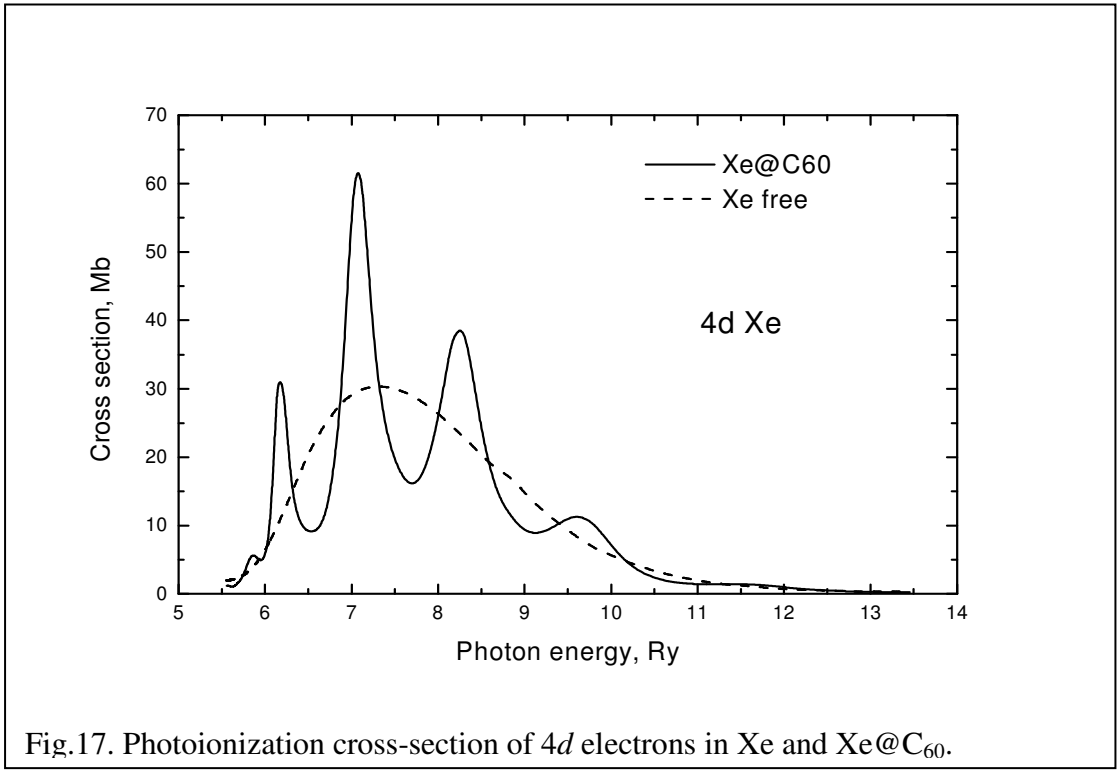

Fig.17. Photoionization cross-section of 4*d* electrons in Xe and Xe@$C_{60}$.

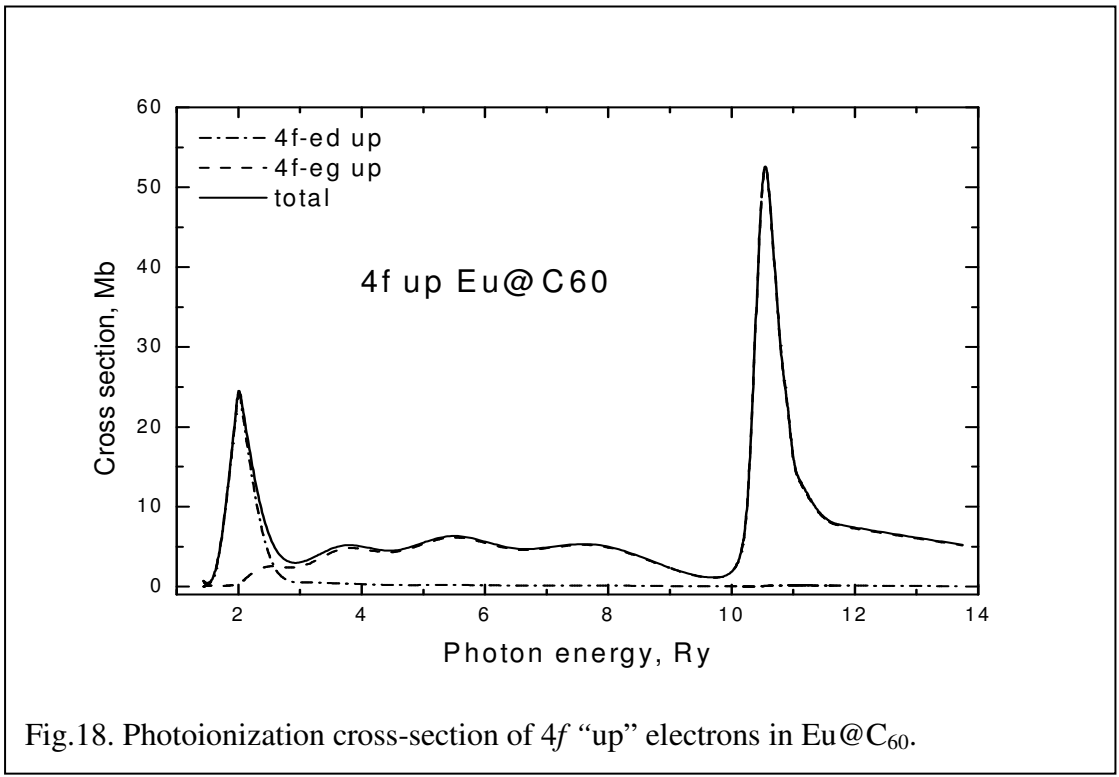

Fig.18. Photoionization cross-section of 4*f* "up" electrons in Eu@$C_{60}$.

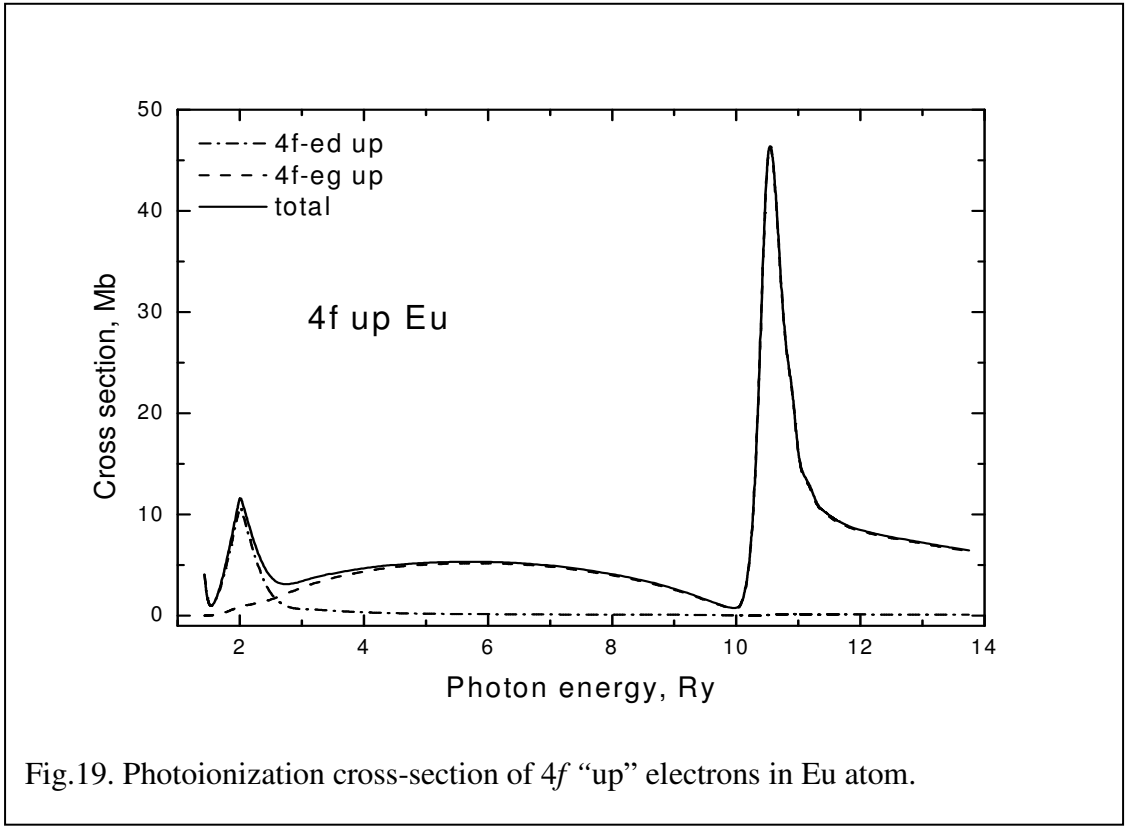

Fig.19. Photoionization cross-section of 4*f* "up" electrons in Eu atom.

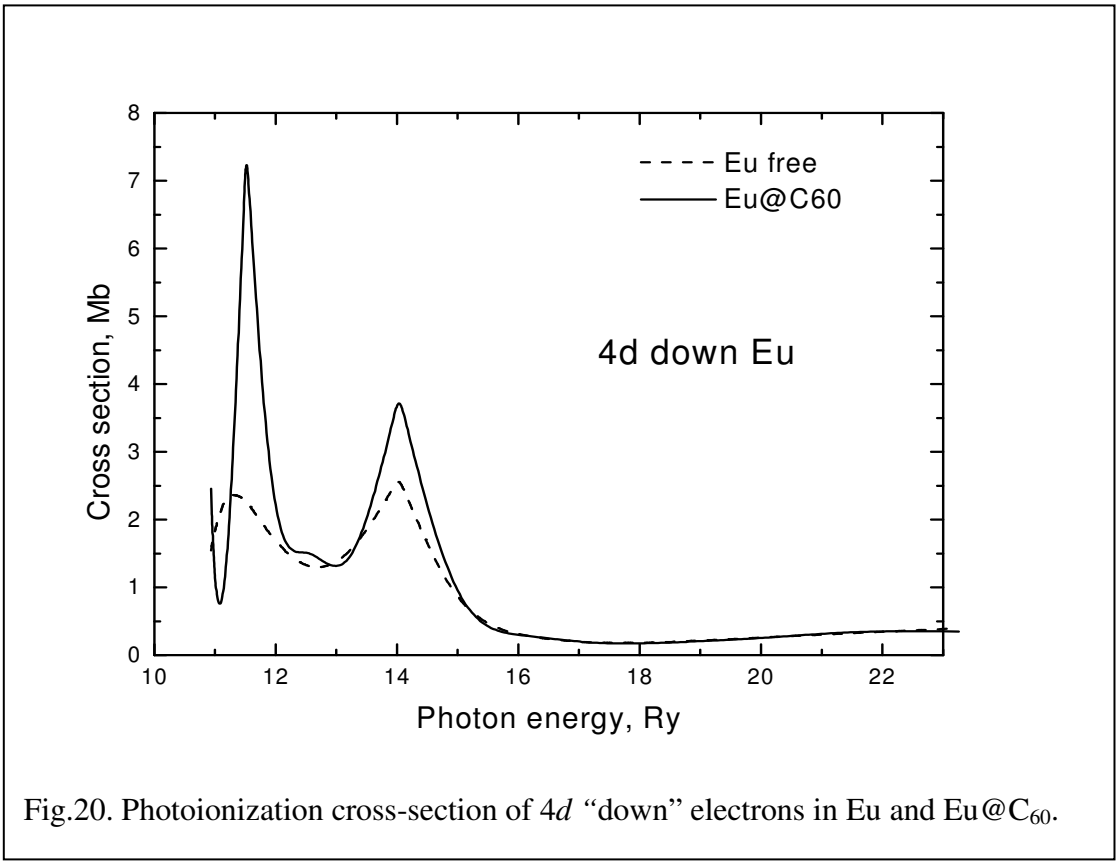

Fig.20. Photoionization cross-section of 4*d* "down" electrons in Eu and Eu@$C_{60}$.

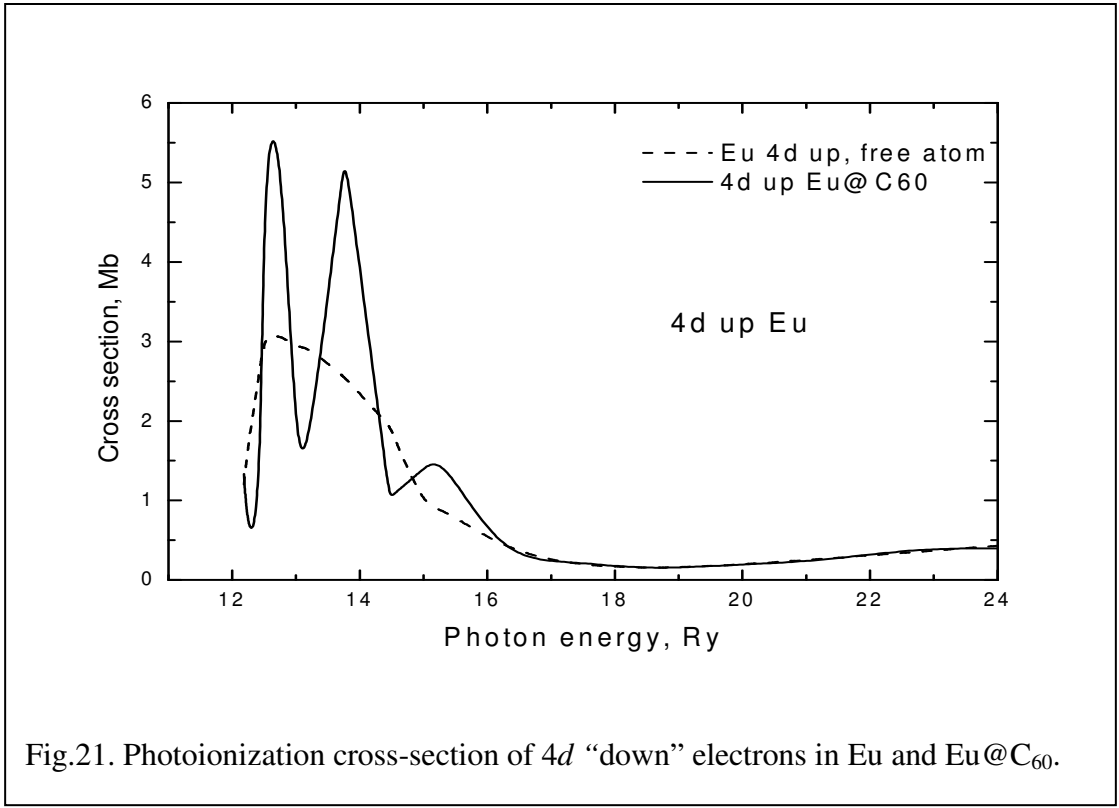

Fig.21. Photoionization cross-section of 4*d* "down" electrons in Eu and Eu@$C_{60}$.

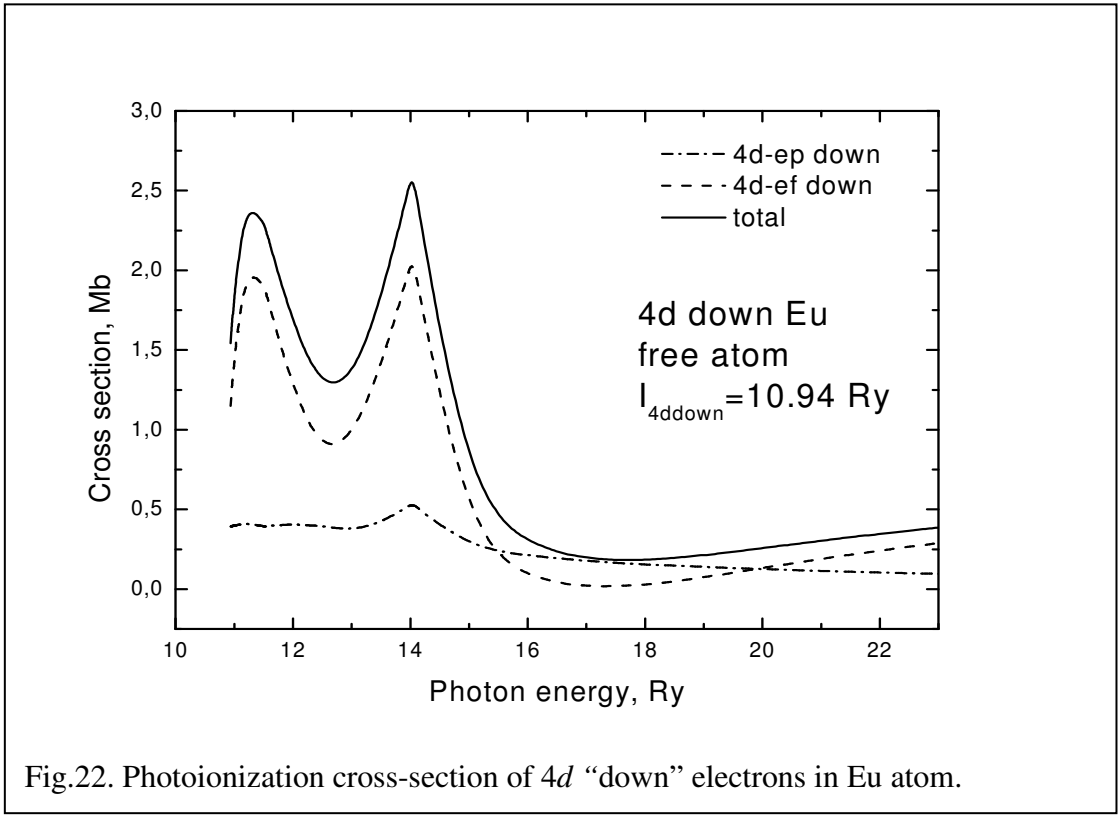

Fig.22. Photoionization cross-section of 4*d* "down" electrons in Eu atom.

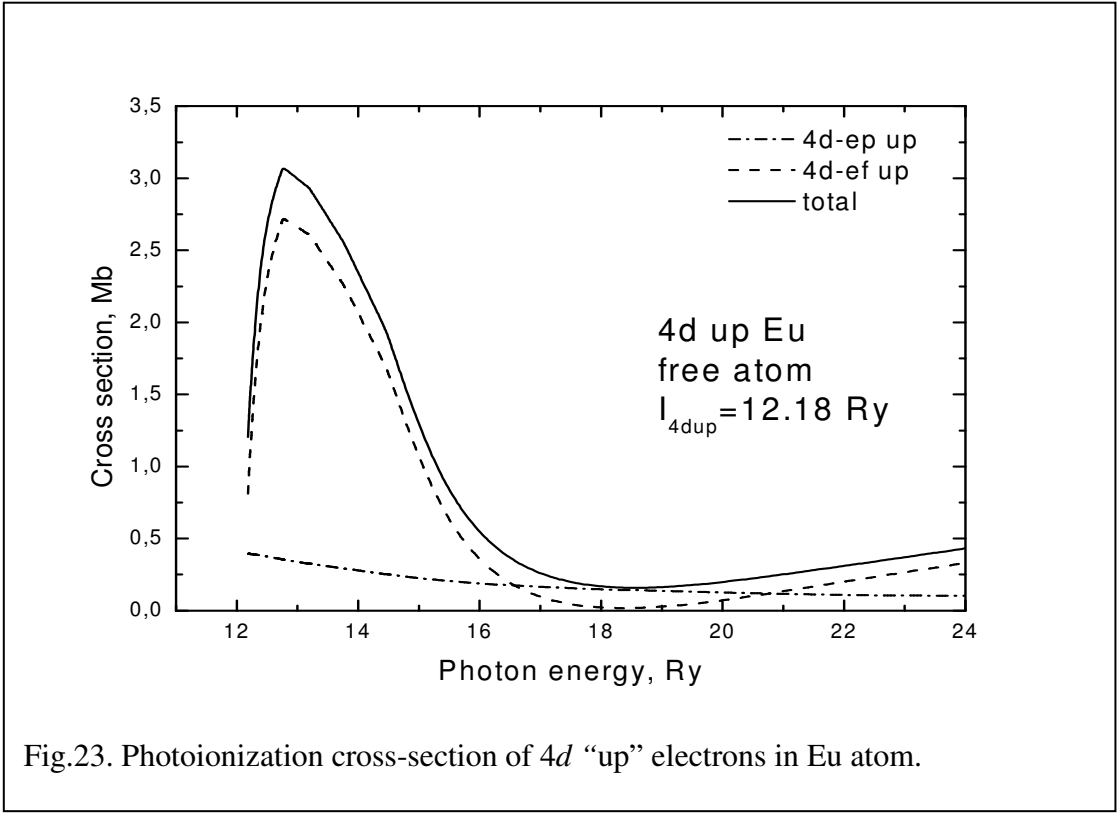

Fig.23. Photoionization cross-section of 4*d* "up" electrons in Eu atom.

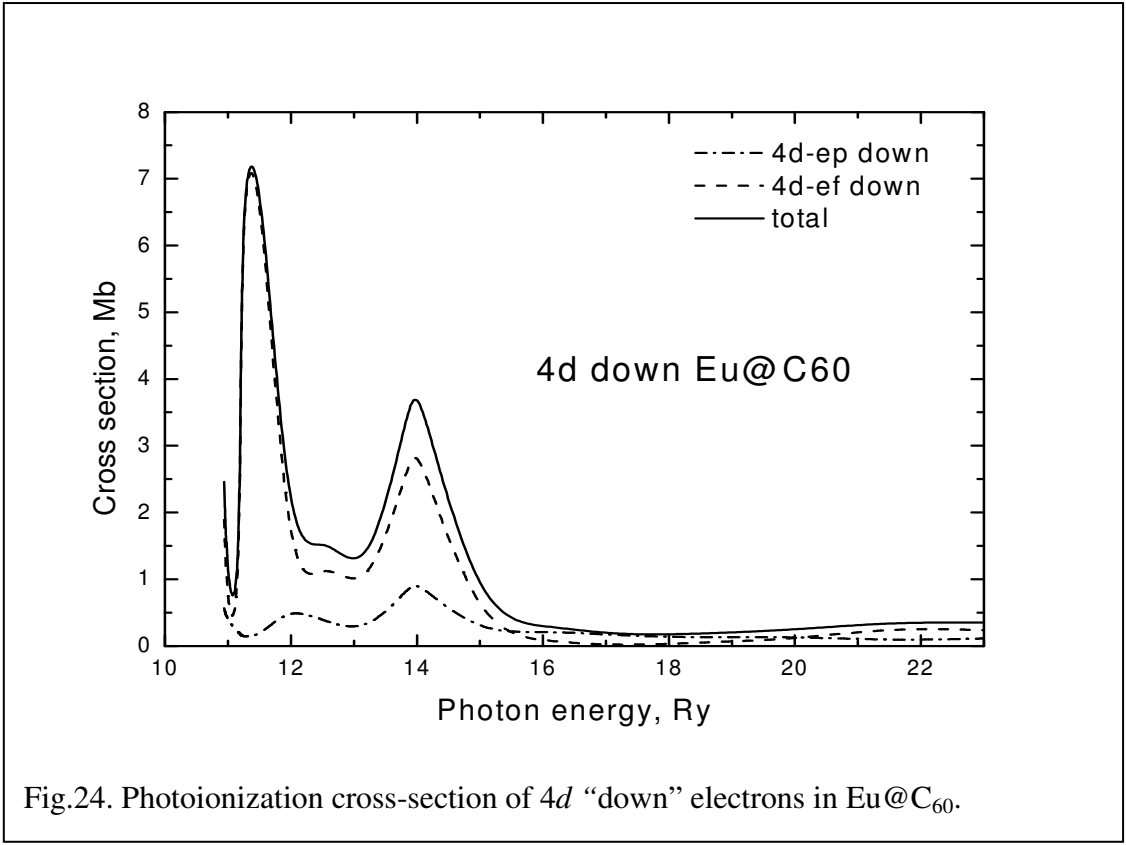

Fig.24. Photoionization cross-section of 4*d* "down" electrons in Eu@$C_{60}$.

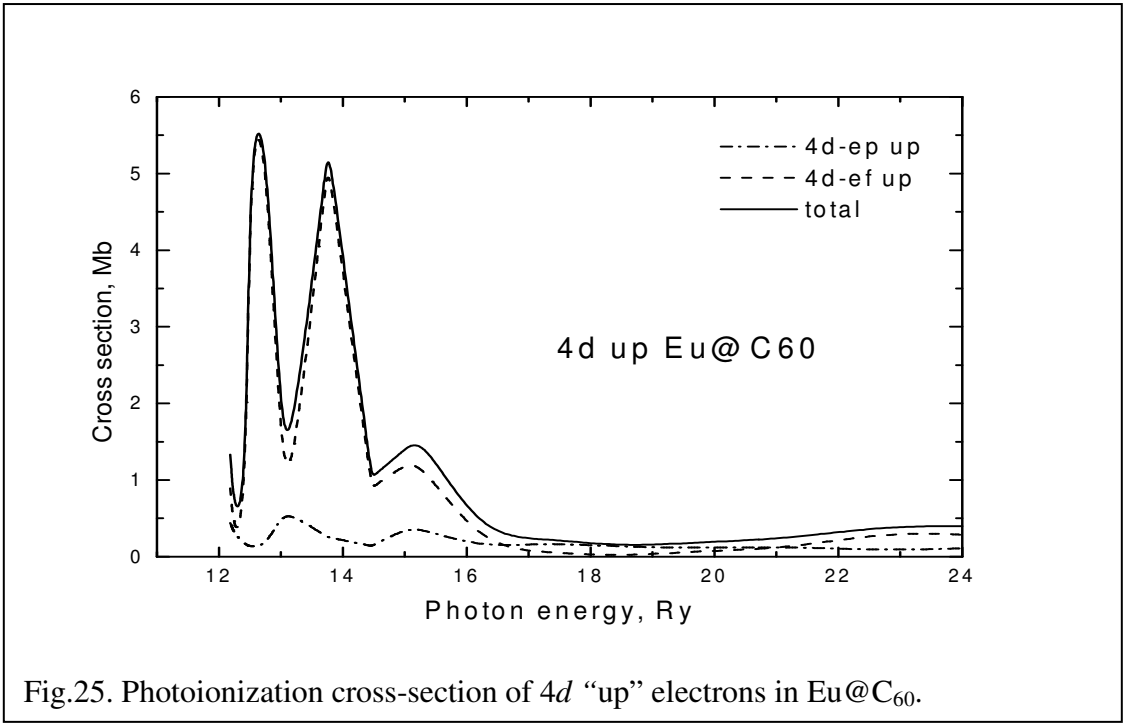

Fig.25. Photoionization cross-section of 4*d* "up" electrons in Eu@$C_{60}$.

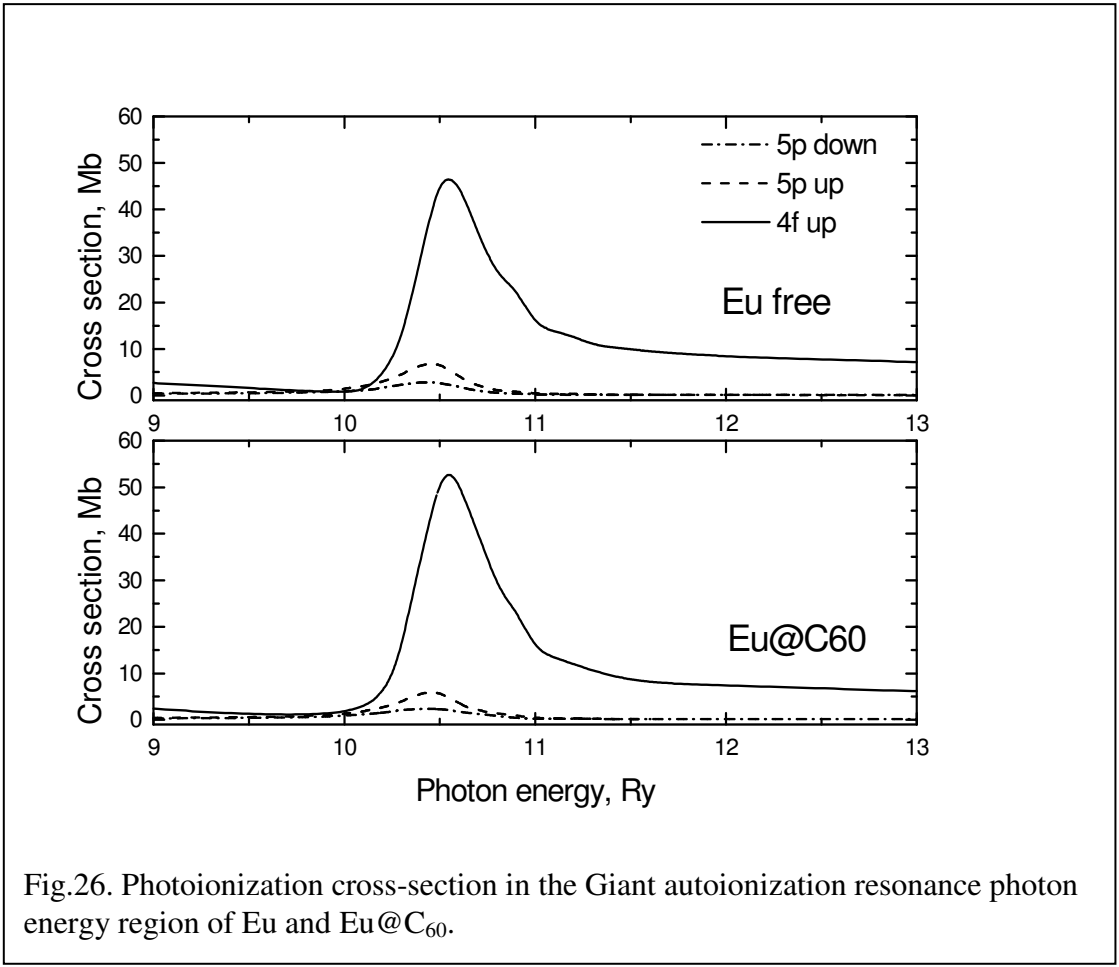

Fig.26. Photoionization cross-section in the Giant autoionization resonance photon energy region of Eu and Eu@$C_{60}$.

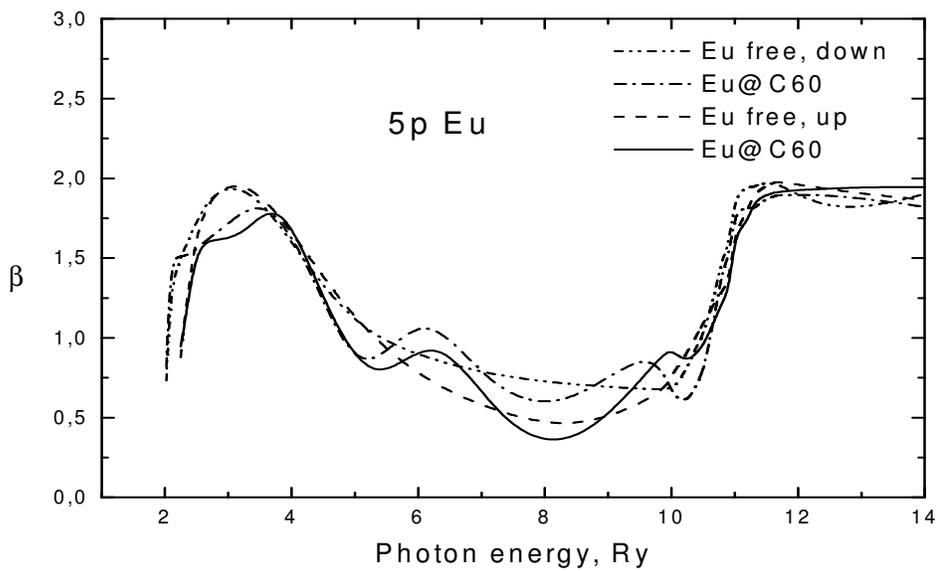

Fig.27. Angular anisotropy parameter of 5p electrons $\beta_{5p}(\omega)$ for Eu and Eu@$C_{60}$.

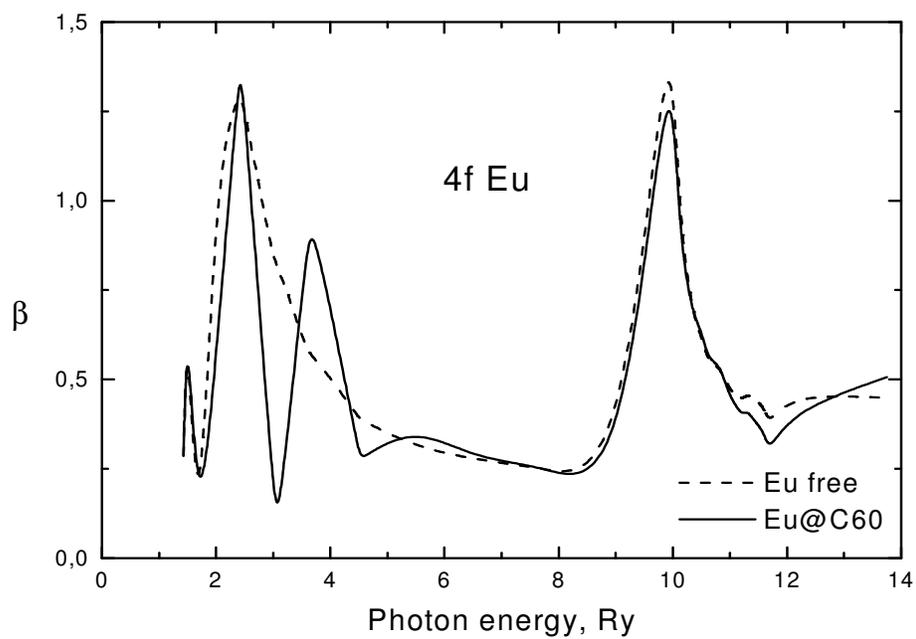

Fig.28. Angular anisotropy parameter of 4f electrons $\beta_{4f}(\omega)$ for Eu and Eu@$C_{60}$.

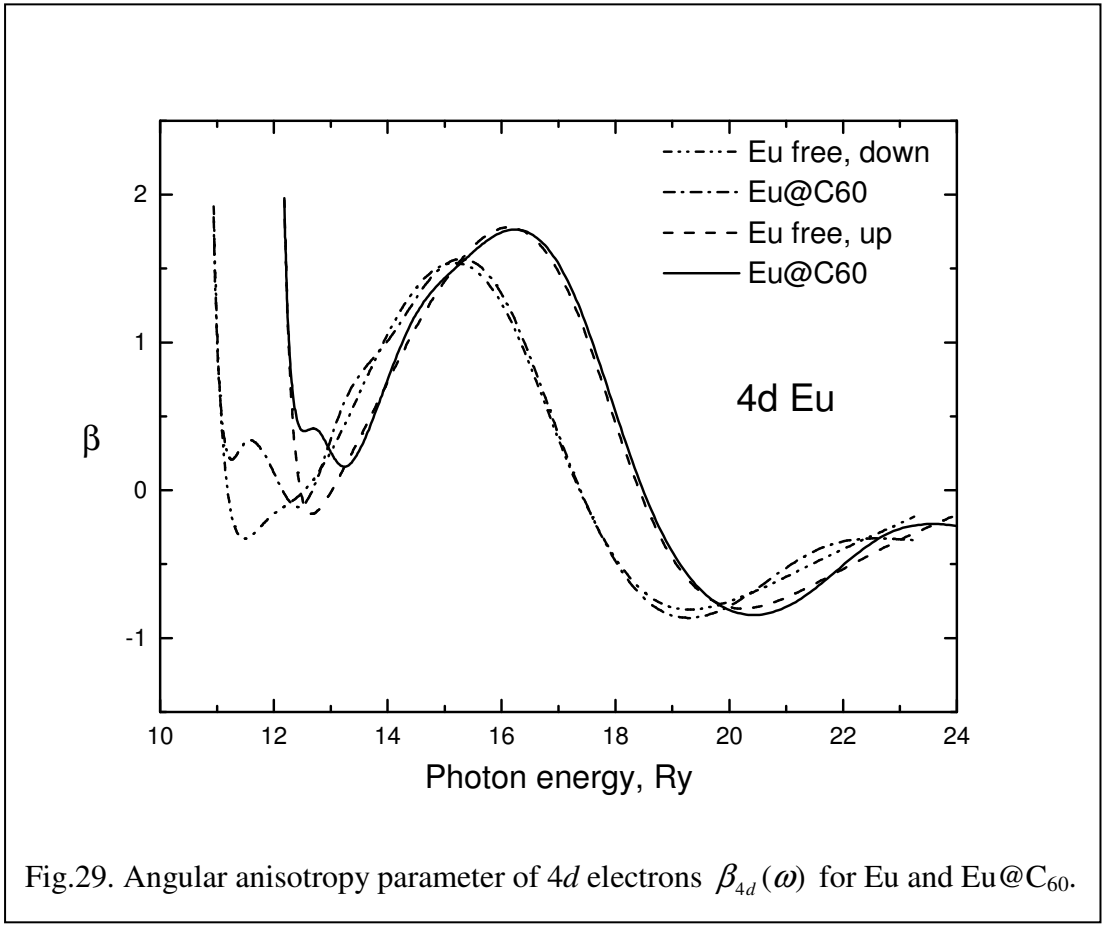

Fig.29. Angular anisotropy parameter of 4$d$ electrons $\beta_{4d}(\omega)$ for Eu and Eu@C$_{60}$.